\documentclass[preprint]{aastex}
\usepackage{psfig,epsfig,flushrt,natbib}

\begin{document}


\title{Clementine Observations of the Zodiacal Light\\
and the Dust Content of the Inner Solar System\vspace*{2ex}}

\author{Joseph M.\ Hahn}
\affil{Lunar and Planetary Institute, 3600 Bay Area Boulevard,
Houston, TX 77058\\
email: hahn@lpi.usra.edu\\
phone: 281--486--2113\\
fax: 281--486--2162\vspace*{2ex}}

\author{Herbert A.\ Zook\altaffilmark{1}}
\affil{NASA Johnson Space Center, Code SN2, 2101 NASA Road 1,
Houston, TX 77058\vspace*{2ex}}

\author{Bonnie Cooper}
\affil{Oceaneering Space Systems, 16665 Space Center Boulevard,
Houston, TX 77058\\
email: bcooper@oss.oceaneering.com\\
phone: 281--228--5332\\
fax: 281--228--5546\vspace*{2ex}}

\and

\author{Bhaskar Sunkara}
\affil{Lunar and Planetary Institute, 3600 Bay Area Boulevard,
Houston, TX 77058\\
email: sunnys@cs.uh.edu\\
phone: 281--480--8304}

\altaffiltext{1}{passed away on March 14, 2001.}

\maketitle

\newpage

\ \\

\begin{center}
Submitted to {\it Icarus} September 27, 2001\\
Accepted March 30, 2002\vspace*{8ex}
\end{center}

\noindent 62 pages\\
2 tables\\
17 figures.\vspace*{8ex}

\noindent keywords: interplanetary dust,
zodiacal light.\vspace*{8ex}

\noindent Running head:\\
The Dust Content of the Inner Solar System\vspace*{8ex}

\noindent Direct editorial correspondence to:\\
Joseph M. Hahn\\
Lunar and Planetary Institute\\
3600 Bay Area Boulevard\\
Houston, TX 77058.

\newpage
\begin{center}
{\bf ABSTRACT}
\end{center}

Using the Moon to occult the Sun, the Clementine spacecraft used
its navigation cameras to map the inner zodiacal light at optical
wavelengths over elongations of
$3\lesssim\epsilon\lesssim30^\circ$ from the Sun. This surface
brightness map is then used to infer
the spatial distribution of interplanetary dust
over heliocentric distances of about 10 solar radii to the orbit
of Venus. The averaged ecliptic surface brightness of the
zodiacal light falls off as
$Z(\epsilon)\propto\epsilon^{-2.45\pm0.05}$
which suggests that the dust cross--sectional density nominally
falls off as $\sigma(r)\propto r^{-1.45\pm0.05}$. The
interplanetary dust also has an albedo of $a\simeq0.1$
that is uncertain by a factor of $\sim2$. Asymmetries of
$\sim10\%$ are seen in directions east--west and north--south of
the Sun, and these may be due the giant planets' secular
gravitational perturbations.

We apply a simple model that attributes the zodiacal
light  as due to three dust populations having distinct
inclination distributions, namely, dust from asteroids and
Jupiter--family comets (JFCs) having characteristic
inclinations of $i\sim7^\circ$, dust from Halley--type comets
having $i\sim33^\circ$,
and an isotropic cloud of dust from Oort Cloud comets.
The best--fitting scenario indicates that asteroids + JFCs
are the source of about $45\%$ of the
optical dust cross--section seen
in the ecliptic at 1 AU, but that at least $89\%$ of the dust
cross--section enclosed by a 1 AU radius sphere is of a
cometary origin. Each population's radial density variations
can also deviate somewhat from the nominal
$\sigma(r)\propto r^{-1.45}$. When these results are
extrapolated out to the asteroid belt, we find an upper limit on
the mass of the light--reflecting asteroidal dust that is
equivalent to a 12 km asteroid, and a similar extrapolation of
the isotropic dust cloud
out to Oort Cloud distances yields a mass
equivalent to a 30 km comet, although the latter mass is
uncertain by orders of magnitude.

\newpage
\section{Introduction}
\label{introduction}

Interplanetary dust is of considerable interest since these
grains represent samples of small bodies that formed in remote
niches throughout the solar system.
Dust grains are liberated when the rocky
asteroids collide and when icy comets sublimate during a close
approach to the Sun, and this dust is transported
throughout the solar system by solar radiation forces. If it is
granted that the information carried by this dust is indeed
decipherable, then samples of this dust tell us about the
conditions in various parts of the solar nebula from
which asteroids, comets, and planets subsequently formed.
In particular, asteroidal dust tells us of nebula conditions 
at the boundary between the terrestrial and the giant--planet
zones in the solar nebula. And since the long--period comets
from the Oort Cloud formed amongst the giant planets,
the mineralogy of their dust is indicative of conditions over a
vast swath of the solar nebula between $\sim5$--30 AU.
Information about the outer reaches of the solar nebula is also
carried by dust generated by the shorter--period Jupiter--family
comets that likely formed in the Kuiper Belt beyond $\sim30$ AU.

This dust is also of dynamical interest since the spatial
density of these `trace particles' allows one to
simultaneously assess the relative strengths of Poynting--Robertson
drag (which drives dust sunwards), the planets' gravitational
perturbations (which disturbs dust orbits and
sometime confines dust at resonances), and mutual collisions
(which fragments and destroys dust). However a deeper
understanding of this dust first requires knowledge of the
abundance and spatial distribution of asteroidal and cometary
dust grains, both of which are the subject of this
investigation.

\citet{W55} performed one of the earliest assessments of the
various sources of interplanetary dust.
This analysis is essentially 
a mass--budget that compares the rates at which
comets and asteroids produce dust to the rates at
which collisions and Poynting--Robertson drag destroy dust.
From the very limited data on comets, asteroids, and
interplanetary dust that were available at the time, it was
concluded that at least 90\% of interplanetary dust is of
cometary origin \citep{W55, W67}. This view prevailed 
for the next three decades until
the Infrared Astronomical Satellite (IRAS) discovered the
asteroidal dust bands \citep{Netal84, Detal84}.
These dust bands are clearly produced by
asteroid families, so these
observations demonstrate that asteroids are also significant
contributors to the interplanetary dust complex, with models
indicating that $\sim30$--$40\%$ of the outer
zodiacal light is due to asteroidal dust
\citep{Detal94, Letal95}. Asteroidal and cometary dust have also
been collected in the Earth's stratosphere by U2
aircraft \citep{Betal93}. Atmospheric entry velocities can be
inferred from these grains, and about $80\%$ of the dust 
in this sample have low entry velocities consistent
with asteroidal orbits. However this finding should be regarded
as an upper limit on the true abundance of asteroidal dust
in the ecliptic since ({\it i.}) dust released from
low--inclination Jupiter--family comets (JFCs) can also have low
entry velocities (see Section \ref{sources}),
and ({\it ii.}) the Earth's
gravitational focusing naturally selects for low--velocity dust
from asteroids (and JFCs as well) over dust from the
higher--inclination Halley--type and Oort Cloud comets.

In order to assess the abundances and spatial
distributions of asteroidal and cometary dust in the inner solar
system, the following analyzes images of the zodiacal light
that were acquired by the Clementine spacecraft while in lunar
orbit. Clementine orbited the Moon for about two months in early
1994, ostensibly to study the lunar surface. But a secondary
objective of this mission was to image the inner zodiacal light
using Clementine's wide-angle navigation cameras. While the Sun
was in eclipse behind the Moon, the Clementine star tracker
cameras acquired hundreds of images of the zodiacal light over
elongations that span the orbit of Venus down to about 10 solar
radii.  As this is the first scientific application of a star
tracker camera, the instrument and its optics are described in
detail in Section \ref{camera} and Appendix \ref{appendixA}.
Section \ref{observations} and Appendix \ref{appendixB}
describe the observations and data reductions, also in some
detail due to several artifacts present in the data.
However the reader uninterested in these particulars can skip
directly to Section \ref{model} where the interplanetary dust
model is described and applied. Results are then summarized in
Section \ref{summary}.

\section{The Star Tracker Camera}
\label{camera}

The zodiacal light images studied here were acquired by
the Clementine spacecraft's two star
tracker cameras. A star tracker is a simple, light--weight,
low--power camera designed to acquire wide--angle CCD images of
star fields. The spacecraft's two star tracker cameras are
designated A and B, and nearly all of the data examined
here were acquired by star tracker B.
The principle purpose of the star tracker is to aid 
spacecraft navigation; by comparing the observed star fields
to an onboard star atlas, the spacecraft can continuously
monitor its orientation. It should be noted that a high--quality
photometric imaging capability was {\sl not} a design criterion
for this camera. Nonetheless, our close inspection of the data
shows that this camera can be used to obtain high--fidelity
images once a number of instrumental artifacts are
removed from the data. These data reductions
are described in detail in Section \ref{observations}.

A simplified schematic of the star tracker optics is shown in
Fig.\ \ref{optics}, and a more detailed description of the
instrumentation may be found in \citet{Letal91} and
\citet{Ketal95}. The camera's principle components are a
spherical lens, a fiber optic, and a CCD detector.
The focal point of this lens is at its center.
As Fig.\ \ref{optics} shows, incident light entering
the lens from the left forms an image at the opposite side
of the lens, and the fiber optic pipes this light to the CCD.
Appendix \ref{appendixA} shows how to map the CCD's $(x,y)$
coordinates for every pixel into equatorial and ecliptic
coordinates.

\begin{figure}[t]
\psfig{figure=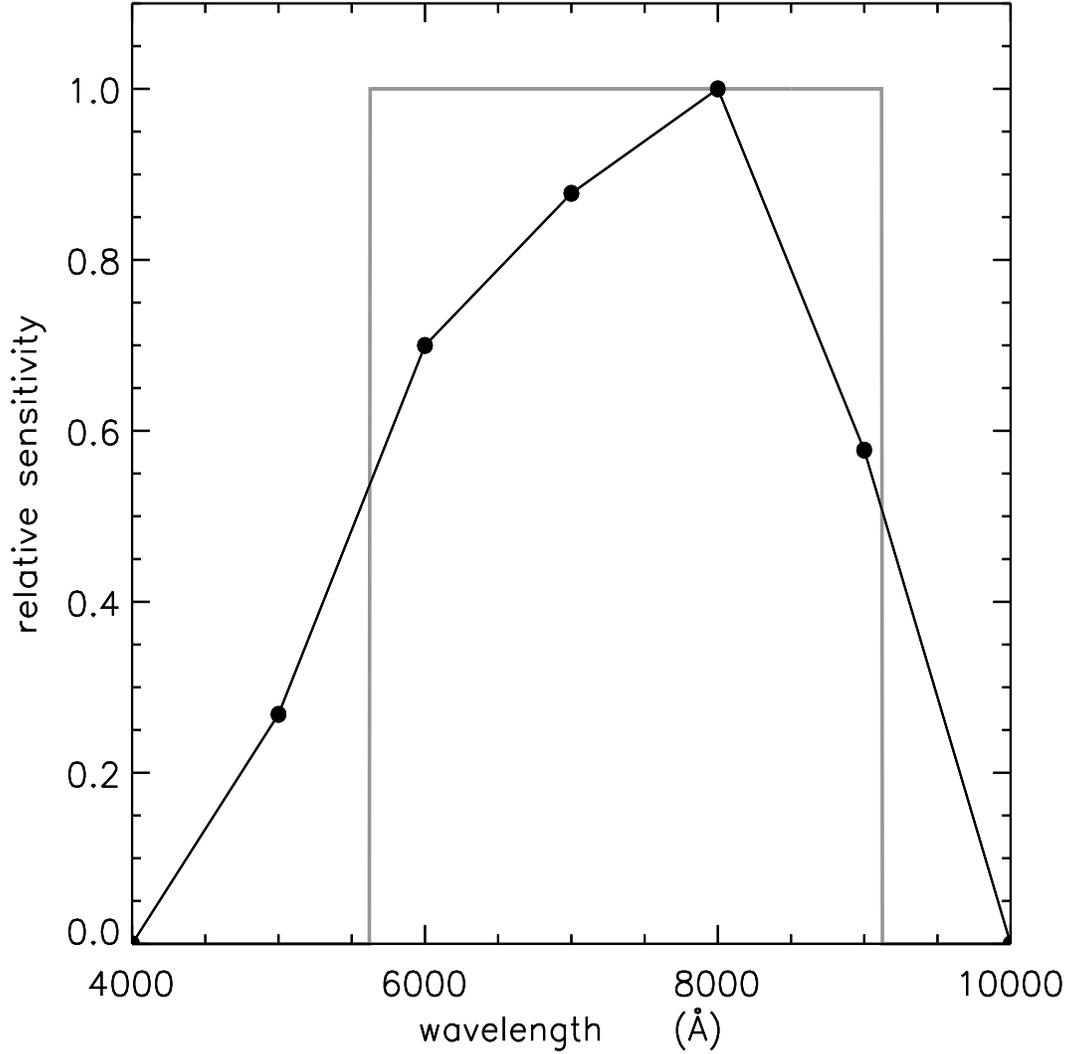,height=6.0in,width=6.0in}
\caption{The dots indicate the relative sensitivity of the star
tracker camera to light of discrete wavelengths of
$\lambda=5000$ to 9000 \AA\  sampled over 1000 \AA\ intervals;
these data are provided by J.\ F.\ Kordas
(private communication). The dark curve simply
connects the dots and also assumes that the camera sensitivity
is zero at $\lambda=4000$ and $\lambda=10000$ \AA.
The grey box is an equivalent square bandpass.}
\label{sensitivity}
\end{figure}

The CCD detector is a Thomson TH7883 array of
$384\times576$ pixels. The camera's angular field of view is
$29.3^\circ\times44.6^\circ$ and the plate--scale at the
optical axis is 0.0756 degrees/pixel. This camera's
point spread function has a full width at half maximum of about
2 pixels $\simeq0.15^\circ$. No filter was used in this
camera. The detector has a peak quantum efficiency of about
$45\%$ at a wavelength of
$\lambda\simeq8000$ \AA\ \citep{Letal91}.
Figure \ref{sensitivity} gives the camera's
relative instrumental response and shows that
the camera is sensitive to wavelengths of
$5000\lesssim\lambda\lesssim9000$ \AA. Figure \ref{sensitivity}
also shows the equivalent square bandpass (e.g., one having the
same area under the curve as the observed instrumental response)
that has a spectral width of $\Delta\lambda=3490$ \AA\ and
a mean wavelength $\bar{\lambda}$=7370 \AA. 

It should be noted that this camera also has two serious
handicaps. The first is that the 8--bit CCD has a dynamic range
of only 256. However the star
tracker images were acquired at exposures that differ by up to a
factor of 14, so the total dynamical range of the zodiacal map
produced here is $14\times256\simeq3600$. Another problem is the
absence of a shutter in the camera which results in the
CCD being continuously exposed as the array is read out along
the detector's columns. This tends to redistribute the flux from
all sources along the CCD columns. However as Section
\ref{observations} shows, this effect is reversible,
and `destreaked' data may be recovered from the raw data
itself.

\section{Observations and Data Reductions}
\label{observations}

\begin{figure}[t]
\psfig{figure=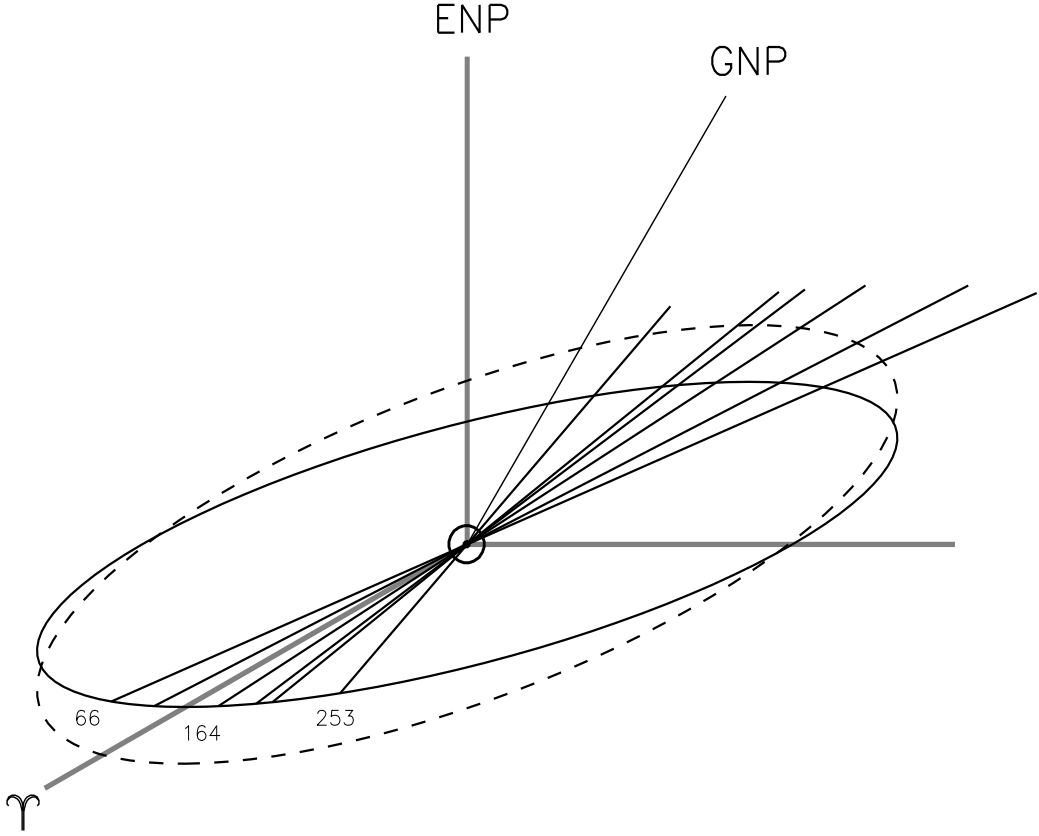,height=6.0in,width=6.0in}
\caption{The narrow lines show the heliocentric ecliptic
longitudes of the spacecraft's
lines--of--sight through the inner zodiacal light, a few of
which are labeled by their orbit number (see Table \ref{orbits}).
Longitudes are measured counter--clockwise from the direction of
the vernal equinox $\gamma$.
The dark ellipse is Earth's orbit about the Sun
$\odot$ with ENP indicating the ecliptic north pole.
The dashed circle represents the $i=3^\circ$ tilt of the
zodiacal light's midplane which has a longitude of ascending node
$\Omega=87^\circ$ \citep{LHRP80}.
The galactic north pole GNP is also
indicated, and it has an heliocentric ecliptic longitude
$\lambda_g=180.0^\circ$ and latitude $\beta_g=29.8^\circ$.}
\label{longitudes}
\end{figure}

The Clementine spacecraft was in an elliptical polar orbit about
the Moon from February 22 through May 4 of 1994, after which it
left the Moon for an encounter with the near--Earth
asteroid Geographos and was subsequently lost
due to a software failure. But during the final
six weeks in orbit about the Moon, the star tracker cameras
repeatedly imaged the inner zodiacal light while the Moon
occulted the Sun. Numerous images were acquired during
an orbit about the Moon, either just prior to sunrise or just
after sunset. Each batch of images are
identified by an orbit number that is simply the number of
lunar orbits that Clementine had achieved to date. Because
these observations were acquired during a six week interval,
the longitude of the camera's line--of--sight to the Sun
changed considerably due to the heliocentric motion of the
Earth--Moon system. Table \ref{orbits} lists orbit numbers,
observation dates, each observation's heliocentric ecliptic
longitude, and total exposure times for the subset of the data
that are examined here. The camera's
lines of sight through the ecliptic during different orbits
are also shown in Fig.\ \ref{longitudes}.

All of the raw Clementine data examined here are archived at the
National Space Science
\mbox{Data Center and may be obtained at the URL}\linebreak
\mbox{http://nssdc.gsfc.nasa.gov/planetary/lunar/clementine1.html},
and the flatfield used to process the star tracker images is
available from the authors.

The star tracker camera usually acquired a sequence of about
40 or so images during each orbit of the spacecraft.
The exposure times for every image acquired during a sequence
usually cycled between 0.05, 0.1, 0.2, 0.4, and 0.7 seconds.
By cycling the exposure times in this manner, the camera's
dynamic range was increased by a factor of 14 and difficulties
due to image--saturation in the brighter parts of the zodiacal
light were mitigated. A typical raw image is shown in
Fig.\ \ref{raw}A, which is a 0.4 sec exposure acquired
during orbit 193. Although the Sun is well behind the
lunar limb, the Moon is partly illuminated by sunlight
reflected by an Earth that is outside of the field of view.
The bright object left of the Moon is a saturated Venus.
In fact, Venus is so bright that the signal accumulating in the
pixel at Venus has bled into the nearest 10 or so pixels.

Note also the bright streak at Venus in Fig.\ \ref{raw}A as well
as a broader but dimmer streak running through the core of the
zodiacal light. These streaks are
a consequence of reading the CCD array in a shutterless camera.
The camera electronics reads the CCD array by shifting the
contents of every pixel down along the CCD's columns
that, in Fig.\ \ref{raw}, run left--to--right.
As a row of ``logical'' pixels shifts
off the bottom of the array, their values are recorded,
zeroed, and then that row shifts back into the CCD's top row.
However the CCD is always exposed during this process,
so reading out the CCD array causes every
logical pixel to receive additional signal from all parts
of the sky that subtend that pixel's CCD column. This results in
an image that appears to have streaks running along the columns
(see Fig.\ \ref{raw}A). Nonetheless, the time to transfer the
contents of one pixel to the next, $94.4\ \mu$sec
\citep{Ketal95}, is a fixed quantity, so the intensity of each
column's streak can be inferred and removed from the raw data
itself using the destreaking algorithm given in \citet{Zetal97}.
However this algorithm fails whenever a CCD column contains one
or more saturated pixels. In this case, the pixels along the
entire column are flagged and are subsequently ignored for the
remainder of the analysis. Pixels polluted by
Venus also disturb the destreaking algorithm, so they are first
replaced with a local average of the zodiacal light prior to
destreaking and are subsequently discarded (see Fig.\
\ref{raw}B).

In principle the dark current should be subtracted before
destreaking an image. However the lack of a shutter makes it
impossible to directly measure the dark current from these
streaked images. Nonetheless, images acquired during orbit 66
show a very dark lunar surface that is shadowed from both the
Sun and the Earth, so the flux observed in lunar surface
provides a first estimate of the dark current. An aperture is
placed on the dark portion of the Moon and an initial estimate
of the dark current $f$ is obtained for every image in the
sequence. This current is subtracted from each image
which is then destreaked. The residual flux $\delta f$ in that
aperture is then examined, and automated software then revises
the estimated dark current $f$ appropriately and this cycle
repeats until $\delta f$ has relaxed to zero.  A similar
algorithm is also used to subtract the dark current from
all of the other images acquired during different orbits. However
these images generally show a lunar surface that is either
slightly or wholly illuminated by earthshine, so the flux
measured in the lunar aperture represents the dark current
plus a nonzero offset $\delta f$. In this case, the above
algorithm iteratively subtracts the dark current $f$ and
destreaks each image until a {\sl predetermined} residual
flux $\delta f>0$ is achieved. The value for $\delta f$
appropriate for each image--sequence is determined later by
comparing images that overlap the orbit 66 field; see below.

The construction of the star tracker flatfield is described in
Appendix \ref{appendixB}. Each image is flatfielded and
pixels that subtend the Moon are flagged and discarded.
Next, small shifts to the images are applied as needed
so that the stars seen in an image--sequence appear stationary.
A single averaged ``master''
image is then formed from the image--sequence
using only the good pixels that were not previously flagged as
bad. Figure \ref{raw}B shows the master image for orbit 193; the
data--gaps correspond to the Moon
as well as pixels polluted by Venus.

Two additional faint artifacts become evident upon
close inspection of this longer--exposure image. The first is
that every eighth column (which runs left--right in Figure
\ref{raw}B) is slightly darker
than its neighbors. It is only evident at the outer edges of the
images where the zodiacal light is quite faint, and its effect
is barely discernible in Fig.\ \ref{raw}B. This is
probably due to a slight inhomogeneity in the dark--current
across the CCD. However this faint striping is of little
consequence since its magnitude is comparable to the noise in
the data. Another faint dark stripe can also be seen in the rows
that subtend Venus; its magnitude is roughly twice the
pixel--to--pixel noise in the image. The cause of this stripe is
unknown, and it is only seen in images containing
a deeply saturated Venus.

Figure \ref{raw}C shows a textbook quality image of the
zodiacal light which has the Moon and Venus pasted back in.

The camera's plate--scale and its pointing
are determined from the handful of
bright field stars that are identified in each of the master
images; see Appendix \ref{appendixA} for details.
Using each star's observed $(x,y)$ coordinates,
their known equatorial coordinates $(\alpha,\delta)$,
and Eqns. (\ref{xy}) and (\ref{Phi_01}), a plate--scale of
$p=0.0756$ degrees/pixel is obtained. The lens coordinates
$(\theta,\phi)$ for each star are then computed (see Fig.\
\ref{optics} and Eqns.\ \ref{xy}),
and Eqns.\ (\ref{a_o,d_o}) are
solved for $(\alpha_o,\delta_o)$, which are the equatorial
coordinates for the pixel at the camera's optical
axis, and $\tau$, which is the angle between the CCD's
$y$ axis and equatorial north. With these quantities
known, equations (\ref{xy}) and (\ref{a_o,d_o}) can now be used
to compute equatorial coordinates for every pixel in each master
image, and Eqs.\ (\ref{lat-long}) are used to rotate these
coordinates into geocentric ecliptic longitude and latitude
$(\Lambda,\Theta)$, as well as the longitude of each pixel
relative to the Sun, $\Lambda-\Lambda_\odot$.

With the pointing for every master image known, it is now
possible to determine the unknown offsets $\delta f$
for the remaining master images that have not yet had their
dark current properly subtracted. By examining those fields that
overlap the orbit 66 field, it is straightforward to estimate the
small offsets $\delta f$ that yield a mutually consistent
surface brightness in the overlapping regions. This process is
then repeated for the remaining adjacent fields until all
offsets for all images have been determined. With these new
offsets in hand, the entire data--reduction cycle
(dark current subtraction,
destreaking, flatfielding, and offset determination) is repeated
until no further changes in the offsets are required.

\begin{figure}
\epsscale{1.0}
\vspace*{-15ex}\plotone{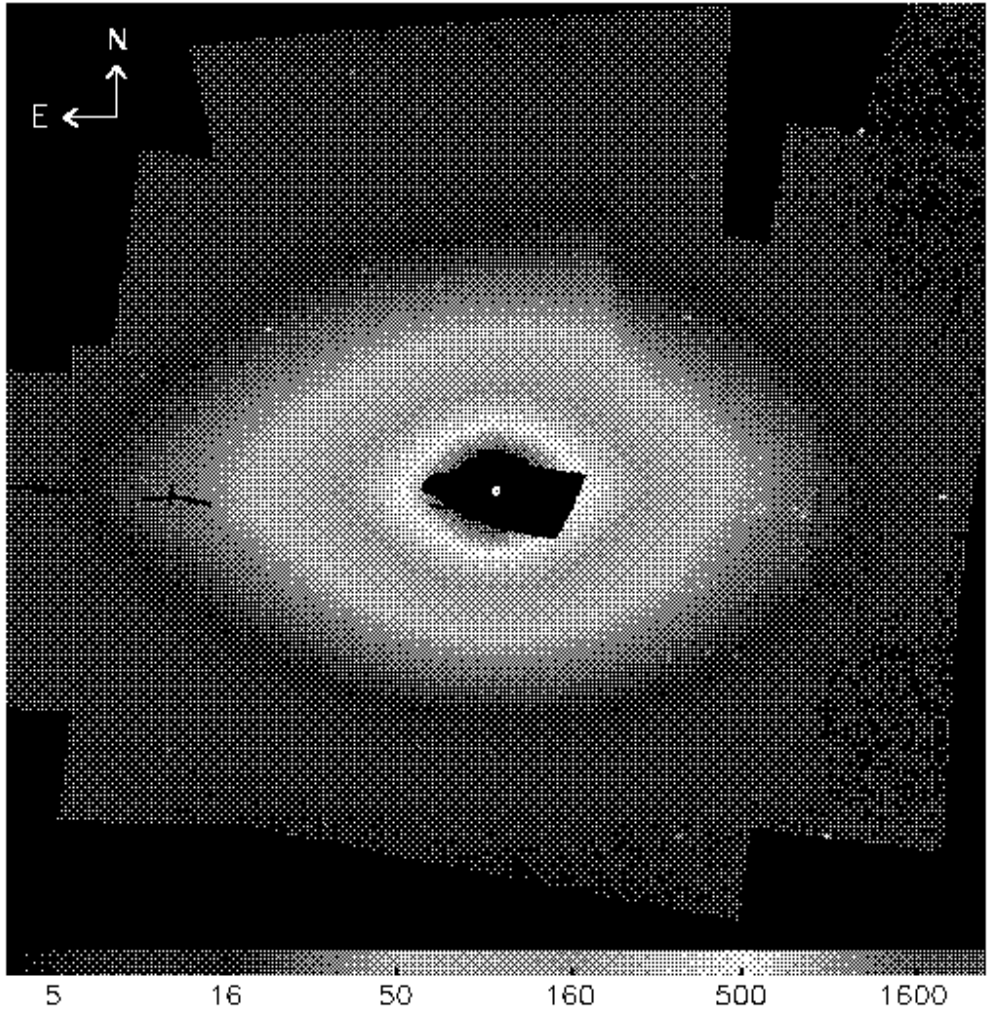}
\vspace*{-15ex}\figcaption{A mosaic of seven fields of the inner zodiacal light
observed by the Clementine star tracker camera. The colorbar
indicates surface brightness in units of $10^{-13}$B$_\odot$.
Ecliptic north and east are up and left in this
mercator projection, and the field of view is $60\times60^\circ$.
Black indicates gaps in the data, and the Sun is drawn to scale
at the center of the mosaic. Regions beyond
$\Phi\sim10^\circ$ northwest of the Sun are polluted by
scattered light, and the ``dimple'' $20^\circ$ east of the Sun
is a lower signal/noise patch that was polluted by Venus.
\label{mosaic}}
\end{figure}

The final step is to merge all of the master images into a single
wide--angle mosaic of the inner zodiacal light. This results in
the $60^\circ\times60^\circ$ mosaic shown in Fig.\ \ref{mosaic}.
This image is formed by mapping the intensity of every good
pixel in all of the master images into the corresponding
$2\times2$ box of pixels in the mosaic, which smooths
the mosaic over an angular scale of $2p=0.15^\circ$.
Also recall that these images were acquired over a six
week interval, so some stars are seen more than once as they
drift to the right with time due to the heliocentric motion of
the spacecraft. Consequently, several planets are also seen at
multiple longitudes: Saturn is barely
discernible at about $11^\circ$ west of the Sun,
Mars appears at $16^\circ$ and $18^\circ$ west, Saturn
again at $19^\circ$, and Mercury at $27^\circ$ west.
If the saturated Venus were not  already clipped from these
images, it would inhabit the data--gap
at $20^\circ$ east of the Sun,

The observed intensity of $\beta$ Hydri is used to calibrate
these data. This G2IV star is the only bright object
in these images having a solar--type spectrum. 
This star has $B-V$ and $U-B$ colors that are solar to
within 0.05 magnitudes, has an apparent $V$ magnitude of
$m_\star=2.80$, and has an instrumental intensity of
$I_\star=770\pm60$ counts/sec. Note that this intensity is
obtained from images that are flatfielded using the
``point--source'' flatfield that is described in
Appendix \ref{appendixB}. One common brightness unit in
zodiacal light observations is
the mean solar brightness $B_\odot=I_\odot/\Omega_\odot$,
which is the intensity of the Sun $I_\odot$ divided
by the solid angle of the Sun $\Omega_\odot=0.223$ deg$^2$.
Since $I_\star=I_\odot 10^{-0.4(m_\star-m_\odot)}$ for a
solar--type star, it follows that
\begin{equation}
\mbox{1 count/sec/pixel}=10^{-0.4(m_\star-m_\odot)}
\frac{\Omega_\sun}{\Omega_p}
\left(\frac{I_\star}{\mbox{counts/sec}}\right)^{-1}B_\odot
\end{equation}
where $m_\odot=-26.78$ is the apparent visible magnitude of the
Sun and $\Omega_p=p^2=5.72\times10^{-3}$ deg$^2$ is the solid
angle of a pixel at the optical axis.
Another common unit is $S10_\odot=4.33\times10^{-16}B_\odot$,
which is the intensity of a tenth magnitude solar type star
distributed over a square degree.
Thus $1\mbox{ count/sec/pixel}=(7.5\pm0.6)\times10^{-14}B_\odot=
(170\pm10) S10_\odot$. The seemingly large uncertainty of $8\%$
in this calibration constant is due to ({\it a}) $\beta$
Hydri's short exposure time of only 0.6 sec, and ({\it b})
noise in the flatfield---see Appendix \ref{appendixB}.

However the relative uncertainties in the mosaic image,
Fig.\ \ref{mosaic}, vary across of the field due to the
different exposure times of the
various master images (see Table \ref{orbits}). The fields
west of Sun, which were acquired during orbits 66 and 110, had
very short exposure times, so the western side of the mosaic is
considerably noisier than the eastern side. Uncertainties in the
dark current subtraction are $\sim3$
counts/sec $\sim2\times10^{-13}B_\odot\sim500 S10_\odot$.
However this uncertainty is significant only at the outer edges
of Fig.\ \ref{mosaic} where it can be as much as $50\%$ of
the signal there. We also note that nearly all of
the images acquired during orbit 164 had a bright Earth in its
field of view, so these images have considerable amounts of
scattered light in them. This is the field just north of the Sun
in Fig.\ \ref{mosaic}, and this polluted zone
lies at elongations of $\Phi\gtrsim10^\circ$
north and north--northwest of the Sun. This is the only field
acquired by star tracker A for which the flatfield is
unavailable. We have elected to process this field using the
flatfield from star tracker B, and it is included in the mosaic
Fig.\ \ref{mosaic} solely for the purpose of filling an
otherwise large datagap. The light--polluted portion of this
field is not used in the subsequent analysis.

Faint, diffuse background light from the galaxy also
contaminates Fig.\ \ref{mosaic}. However this was minimized by
observing at an epoch when the sunward lines of sight were at
the highest possible galactic latitudes of
$30^\circ\lesssim\beta_g\lesssim90^\circ$
(see Fig.\ \ref{longitudes}). The surface brightness of the
galaxy was measured by Pioneer 10 while at heliocentric
distances $r>2.8$ AU where the zodiacal contribution is
negligible; at latitudes $\beta_g>30^\circ$ the galactic surface
brightness is $Z_g<90\ S10_\odot$ \citep{Letal98} at the
southern edge of Fig.\ \ref{mosaic}, and it decreases to
the north. However this flux is substantially smaller than the
uncertainty in the dark current subtraction and is neglected
here.

The integrated intensity of the light seen in Fig.\ \ref{mosaic}
is $I=4.8\times10^{-8}I_\odot$ which corresponds to a visual
magnitude $m_V=-8.5$. This makes the zodiacal light the second
brightest object in night sky, the first being the full Moon
having $m_V=-12.7$ and the third being Venus with $m_V=-4.6$ at
its brightest.

\section{A Simple Model of the Interplanetary Dust Complex}
\label{model}

In order to extract the gross properties of the observed dust
seen in Fig.\ \ref{mosaic}, a simple model that is quite common
in the literature shall be fitted to these data [{\it c.f.},
\cite{Leinert75}]. The model assumes that the dust density
varies radially as a power--law with heliocentric distance 
and that the cloud is axially symmetric.
This model also assumes that the center of the cloud is at the
Sun and that its midplane is in the ecliptic. Although
none of these assumptions are actually correct in detail,
they are sufficiently good for our purposes. In this case the
spatial density of dust cross--section $\sigma$ can be
written as a function of the heliocentric distance $r$ and
heliocentric ecliptic latitude $\beta$:
\begin{equation}
\sigma(r,\beta)=\sigma_1\left(\frac{r}{r_1}\right)^{-\nu}h(\beta)
\end{equation}
where $r_1=1$ AU is a reference distance,
$\sigma_1=\sigma(r_1,0)$ is the dust cross--section density in
the ecliptic at $r=r_1$, and $h(\beta)$
describes how the dust density falls off with ecliptic latitude.

\begin{figure}
\plotone{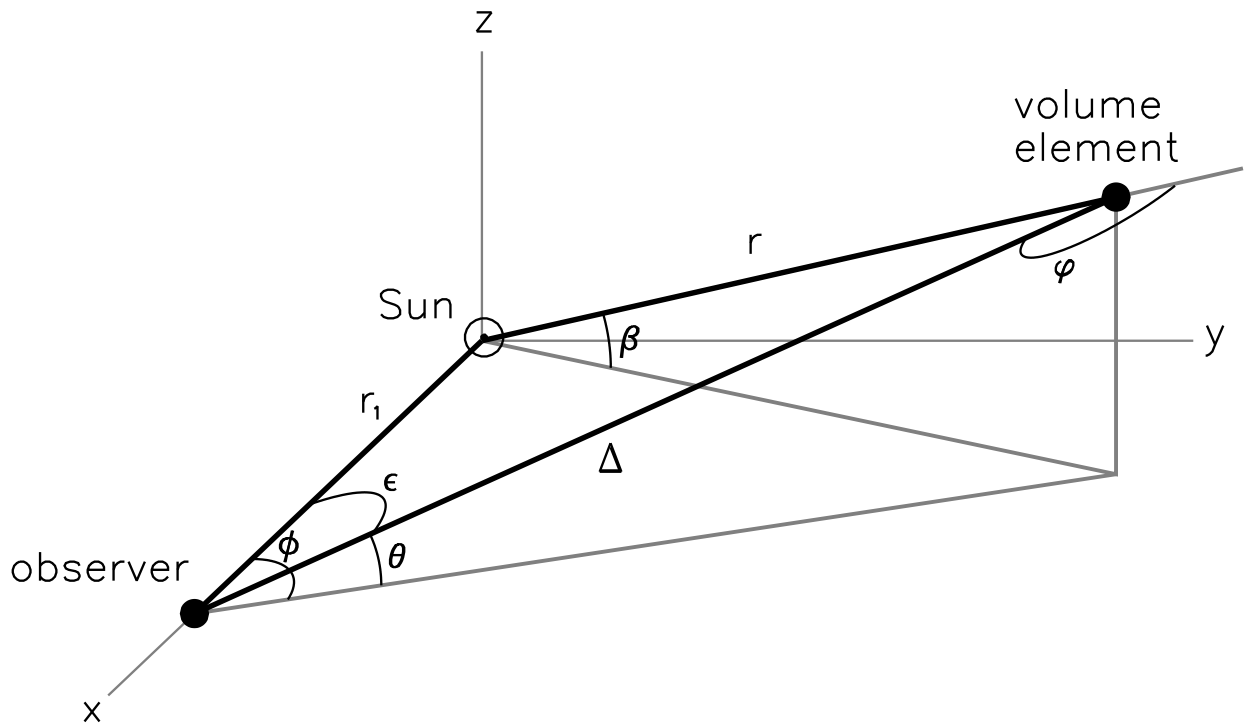}
\figcaption{The viewing geometry for light scattered by a dust
volume element a distance $r$ from the Sun at
heliocentric latitude $\beta$
measured from the ecliptic $x$--$y$ plane. The observer has a
heliocentric distance $r_1$ and the line--of--sight (LOS) to
the volume element a distance $\Delta$ away has a
geocentric latitude
$\theta$, longitude $\phi$, and an elongation angle $\epsilon$
measured from the sunward direction. The scattering angle
$\varphi$ is measured from the LOS to the anti--solar direction
at the volume element.
\label{geometry}}
\end{figure}

The surface brightness of the sunlight reflected by
this dust distribution is proportional to
$\sigma(r,\beta)$ multiplied by a light--scattering function and
integrated along an observer's line--of--sight.
The flux density of sunlight that is reflected
by dust in a small volume element $dV$ is
$dF=\sigma(r,\beta)\Phi(\varphi)(L_\odot/4\pi r^2)dV/\Delta^2$
\citep{Letal79} where $dV=\Omega\Delta^2d\Delta$ and $\Omega$ is
the solid angle of the volume element as seen by an observer a
distance $\Delta$ away; see Fig.\ \ref{geometry} for the
definition of all the geometric quantities used here.
The scattering phase function $\Phi(\varphi)$ is
related to the phase law $\psi(\varphi)$ via
$\Phi(\varphi)=(a/\pi\mbox{ sr})\psi(\varphi)$ where $a$ is
the dust geometric albedo and $\varphi$ is the scattering angle.
Note that this formulation is valid only in the geometric optics
limit, which is indeed the case since the bulk of the dust
cross--section is contributed by grains having sizes
$\sim10$--100 $\mu$m \citep{Getal85}.

The surface brightness of the
zodiacal light is thus $Z=\int dF/\Omega$ integrated over
$0\le\Delta<\infty$. Noting that
$r/r_1=\sin\epsilon/\sin\varphi$ where $\epsilon$ is the
elongation of the line of sight having a
geocentric ecliptic latitude
and longitude $(\theta,\phi)$ relative to the Sun, then
$\cos\epsilon=\cos\phi\cos\theta$,
$\Delta/r_1=\sin(\varphi-\epsilon)/\sin\varphi$,
$d\Delta/r_1=\sin\epsilon d\varphi/\sin^2\varphi$, and so
the surface brightness can be recast as
an integral over the scattering angle $\varphi$
\citep{Aetal67, GD69}:
\begin{equation}
\label{Z}
Z(\theta,\phi)=\frac{a\sigma_1 r_1}{\sin^{\nu+1}\epsilon}
\left(\frac{\Omega_\odot}{\pi\mbox{ sr}}\right)B_\odot
\int_\epsilon^\pi\psi(\varphi)h(\beta(\varphi))
\sin^\nu(\varphi)d\varphi
\end{equation}
where $\Omega_\odot=0.223$ deg$^2=6.80\times10^{-5}$ sr
is the solid angle of the Sun and
$B_\odot=L_\odot/4\pi r_1^2\Omega_\odot$ is the mean surface
brightness of the Sun where $L_\odot$ is the solar luminosity.
Note also that the $\beta$ in Eq.\ \ref{Z} depends on the
scattering angle $\varphi$ through
$\sin\beta=\sin(\varphi-\epsilon)\sin\theta/\sin\epsilon$.

\subsection{Radial variations}
\label{radial}

\begin{figure}
\plotone{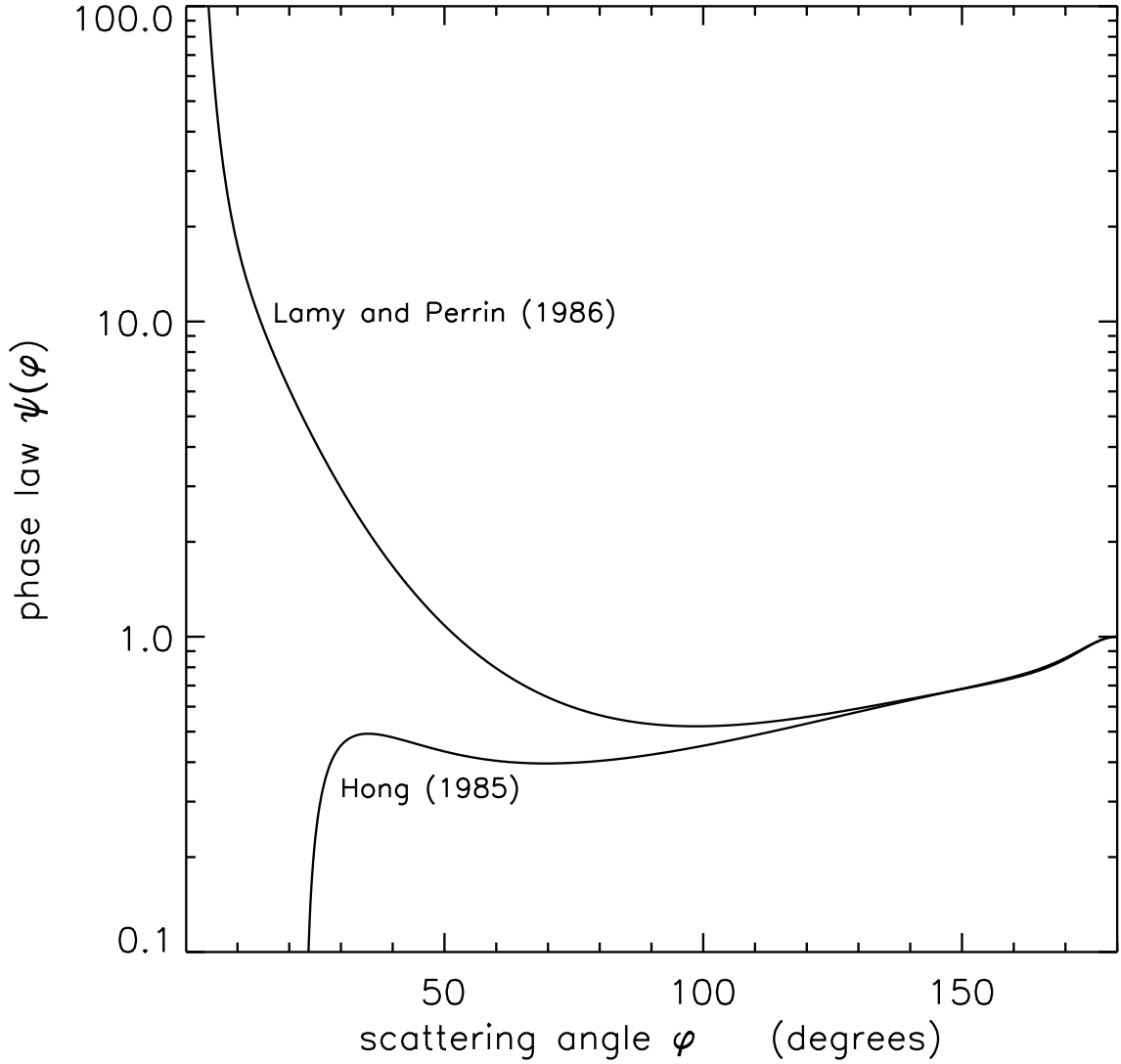}
\figcaption{The upper curve is the phase law obtained from the
nominal volume scattering function of \cite{LP86},
while the lower curve is from \cite{Hong85}. Note that the
phase laws shown here are simply the volume scattering
functions with the dust albedo $a$ and cross sectional density
$\sigma_1$ factored out and normalized to unity
at $\varphi=\pi$ where the phase angle $\pi-\varphi=0$.
Although \cite{Hong85} gives a phase law
for a $\nu=1$ radial power law, we have used his
Eqns.\ (10) and (14) to form the phase law shown here
that is appropriate for a $\nu=1.45$ dust distribution.
\label{phase_fn}}
\end{figure}

For a line of sight in the ecliptic, $h(0)=1$ and
Eq.\ (\ref{Z}) becomes a simple integral over the phase law
$\psi(\varphi)$. Two very different empirical phase laws are
shown in Figure \ref{phase_fn}. The upper curve was constructed
by \citet{LP86}, and it exhibits a very strong forward
scattering peak ({\it i.e.}, $\psi$ diverges as
$\varphi\rightarrow0$) as might occur due to the diffraction of
sunlight by dust larger than a wavelength. The lower phase law
is from \citet{Hong85}; although this law does not show any
forward scattering, it does exhibit
a mild backward scattering peak
at $\varphi\simeq180^\circ$ as is required of
any phase law in order to reproduce the gegenshein. Despite
the very different forms for $\psi(\varphi)$, both phase laws
are very able at reproducing a varied suite of other zodiacal
light measurements that were acquired over a wide range of
elongation angles $\epsilon$ \citep{Hong85, LP86}.
Consequently, our results are remarkably
insensitive to the choice of the phase law. Regardless of
whether one adopts Hong's backscattering phase function or
Lamy and Perrin's forward scattering phase law, a numerical
integration of the Eq.\ (\ref{Z}) yields a surface
brightness $Z(\epsilon)$ that, over our observation interval
$2\lesssim\epsilon\lesssim30^\circ$, is largely indistinct aside
from a numerical factor of $\simeq1.6$. This insensitivity to the
details of $\psi(\varphi)$ is due to the fact that the dominant
contribution to the surface brightness integral is by dust
in the vicinity of $\varphi\simeq90^\circ$, {\it i.e.}, dust
nearer the Sun along the line--of--sight.

This particular behavior also means that
the integral in Eq.\ (\ref{Z}) is quite insensitive to the lower
integration limit for the range of elongations
$\epsilon\lesssim30^\circ$ that are considered here.
In this case the line of sight integral
evaluates to $\simeq0.83$ when the Hong phase law is used
and $\simeq1.3$ when the Lamy and Perrin law is used. 
Eq.\ (\ref{Z}) then simplifies to
\begin{equation}
\label{Zo}
Z(\epsilon)\simeq(2.3\pm0.5)\times10^{-5}
\frac{a\sigma_1 r_1}{\sin^{\nu+1}\epsilon}B_\odot
\end{equation}
where the error in the coefficient indicates the uncertainty in
the phase law. We also note that about
$90\%$ of the light seen in the ecliptic at elongations
$\epsilon\le30^\circ$ is contributed by dust orbiting interior
to $0.6$ AU.

\begin{figure}
\plotone{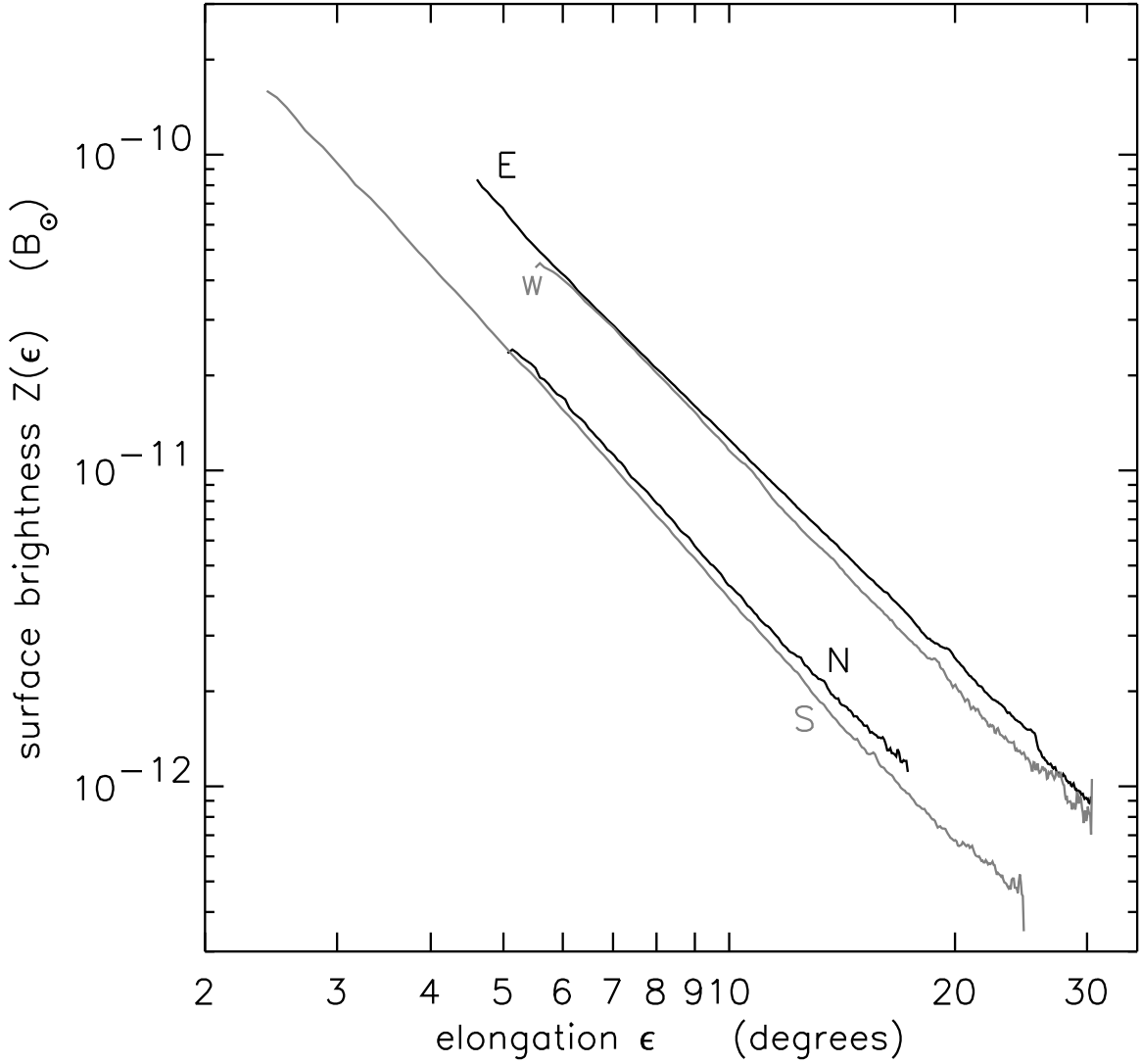}
\figcaption{The surface brightness of the zodiacal light mosaic
of Fig.\ \ref{mosaic} is plotted versus elongation angle
$\epsilon$. Each profile is computed in a $10$ degree--wide
triangular aperture oriented North, South, East, or
West of the Sun. The light--polluted field acquired during orbit
164 (which lies north of the Sun) is discarded before generating
these profiles, as are pixels that subtend data--gaps, bright
stars, or planets.
\label{profiles}}
\end{figure}

The radial power law $\nu$ is now readily obtained from
profiles of the zodiacal light's ecliptic surface brightness.
East--west and north--south profiles are shown in
Fig.\ \ref{profiles}, and a power--law fit to the averaged
east--west profile yields
\begin{equation}
\label{Zobs}
Z(\epsilon)=\frac{(1.7\pm0.2)\times10^{-13}}
{\sin^{2.45\pm0.05}\epsilon}B_\odot
\end{equation}
so a comparison with Eq.\ (\ref{Zo}) shows that
$\nu=1.45\pm0.05$ and $a\sigma_1r_1=(7.4\pm1.8)\times10^{-9}$.
The uncertainty in the former quantity includes
the statistical variations of the data seen in
Fig.\ \ref{profiles} while the latter quantity also includes the
$8\%$ uncertainty in the calibration and the uncertainty due to
the possible choices for the phase law. Note that the power law
$\nu$ reported here is slightly steeper than that inferred from
the data obtained by the Helios spacecraft \citep{Letal81}
and the Cosmic Background Explorer (COBE)
spacecraft \citep{Ketal98},
with the disagreements at the $2\sigma$ level.

\begin{figure}
\epsscale{1.0}
\plotone{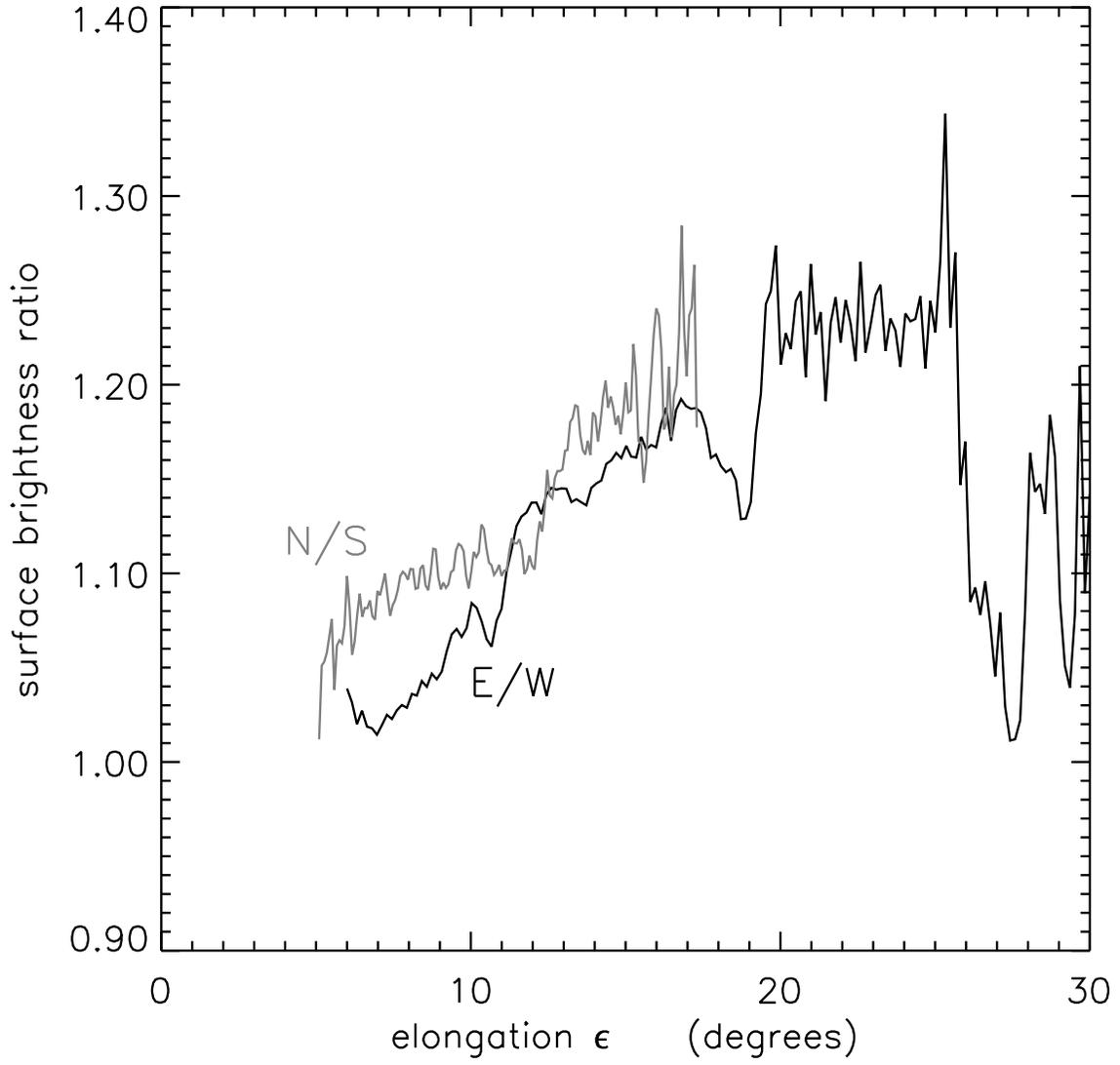}
\figcaption{Ratios of the brightness profiles of Fig.\
\ref{profiles} are plotted versus elongation angle $\epsilon$,
where N/S and E/W indicate the north/south and east/west ratios.
\label{ratios}}
\end{figure}

Figure \ref{profiles} also reveals an asymmetry in the surface
brightness of the zodiacal light north/south of the ecliptic, as
well as an asymmetry east/west of the Sun.
Ratios of the north/south and east/west
surface brightness profiles are plotted
in Fig.\ \ref{ratios} which shows that the northern
hemisphere gets steadily brighter with elongation
relative to the southern hemisphere, as does the eastern ansa
relative to the western ansa.
Asymmetries such as these have been attributed to the giant
planets' secular gravitational perturbations which can organize
the longitudes of the dust grains' perihelia and nodes
\citep{Wetal99}.
The north--south asymmetry seen here is likely the same asymmetry
previously observed by the Helios 1 and 2 which detected a
$i=3^\circ$ tilt between the midplane of the inner
zodiacal light and the ecliptic \citep{LHRP80}. The node
of this symmetry plane has a longitude of
$\Omega=87^\circ$ which, as Fig.\ \ref{longitudes}
shows, is largely perpendicular to the Clementine
lines--of--sight, and this particular viewing geometry
will make one hemisphere slightly brighter than the other.
Gravitational perturbations by giant planets
can also displace the zodiacal cloud's center of light
radially away from the Sun, which results in a phenomenon
known as pericenter glow \citep{Wetal99}; such
perturbations may be responsible for the east--west asymmetry
seen in  Figs.\ \ref{profiles}--\ref{ratios}.

A comment on the Lamy and Perrin volume scattering function is
also in order. \cite{LP86} adopt a volume scattering function
$\Psi$ (which is proportional to the product $a\sigma_1\psi$
used here) that varies
with heliocentric distance as $r^{-\nu_c}$ where $\nu_c$
is chosen so that the dust cross sectional density
$\sigma(r)$ falls off as $r^{-\nu_1}$ where $\nu_1=1$.
Although \cite{LP86} provide excellent arguments to
motivate their approach, their assumption will not be adopted
here, but only because we wish to compare our findings to other
studies of the zodiacal cloud that similarly do not apply this
assumption. However our findings reported below are easily
recalibrated if the Lamy and Perrin volume scattering function
is preferred. If it is assumed that $\Psi$ does indeed vary as
$r^{-\nu_c}$ then this implies that the dust albedo also varies
as the power law $a(r)\equiv a_1(r/r_1)^{-\nu_c}$ where $a_1$ is
the dust albedo at $r=1$ AU. Accounting for this alternate
interpretation thus requires the substitutions
$a\rightarrow a_1$ and $\nu\rightarrow\nu_1+\nu_c$ in
Eqs.\ (\ref{Z}--\ref{Zo}), but this has no substantive effect on
their form. The only significant changes to our findings would
be ({\it i.}) that the albedo quoted in Section \ref{abundances}
should be interpreted as the albedo $a_1$ for dust at 1
AU and that ({\it ii.}) the integrated dust surface densities
and masses given in Eqs.\ (\ref{Sigma1}--\ref{Miso})
need to be reevaluated for the alternate dust density
power--law $\nu_1$. But as long
as $\nu_c$ (which might be as large as $0.45$)
is smaller than unity then these revisions will
change our finding by factors that are also of of order unity.

\subsection{Vertical variations}
\label{vertical}

Estimates of the dust grains' vertical distribution is obtained
by first developing a simple yet plausible model for the dust
inclination distribution $g(i)$. The dust latitude distribution
$h(\beta)$ is then
calculated from this inclination distribution, and
a surface brightness map of the model dust cloud is
generated using the line--of--sight integral Eq.\ (\ref{Z}).
Then by comparing isophotes of the model cloud to the
observations (Fig.\ \ref{mosaic}), the suite of dust models
that are consistent with the observations are readily obtained
below.

\subsubsection{inclination distributions}
\label{inclinations}

Section \ref{sources} will describe the various inclination
distributions $g_j(i)$ for distinct dust--sources {\it i.e.}, the
asteroids, comets, and interstellar sources that are indicated
by the $j$ subscript. The relationship between population $j$'s
latitude distribution $h_j(\beta)$ and its inclination
distribution $g_j(i)$ is
\begin{equation}
\label{g-h}
h_j(\beta)=\int^{\pi/2}_{\beta}\frac{g_j(i)di}
{\sqrt{\sin^2i-\sin^2\beta}}
\end{equation}
\citep{Divari68, D93, B01}. Although Eq.\ (\ref{g-h}) is formally
derived for bodies in circular orbits, it nonetheless provides
reliable results even for bodies in very eccentric orbits
[c.f.\ \cite{B01}]. Note that an
isotropic cloud has an inclination distribution
$g_j(i)=(2/\pi)\sin i$ which results in a latitude distribution
$h_j(\beta)=1$. Accordingly, the total distribution
will be written as a sum over several
possible components, one being an isotropic source with an
{\it iso} subscript and the $N$ other sources having a gaussian
distribution of inclinations:
\begin{equation}
\label{g(i)}
g_j(i)=\frac{2}{\pi}\sin i\times\left\{
\begin{array}{ll}
1 & {j=\mbox{iso}}\\
c_je^{-(i/\sigma_j)^2/2} & \mbox{otherwise}
\end{array}
\right.
\end{equation}
where $\sigma_j$ is the standard deviation of each component's
inclination distribution and $c_j$ is a normalization constant
such that $h_j(0)$ is unity. If each population's
cross--sectional density is assumed to vary as a distinct
power--law $\propto r^{-\nu_j}$, then the zodiacal light's total
surface brightness becomes [see Eqn.\ (\ref{Z})]
\begin{equation}
\label{Z_j}
Z(\theta,\phi)=\sum_jf_jZ_j(\theta,\phi)=a\sigma_1 r_1
\left(\frac{\Omega_\odot}{\pi\mbox{ sr}}\right)B_\odot
\sum_j\frac{f_j}{\sin^{\nu_j+1}\epsilon}
\int_\epsilon^\pi\psi(\varphi)h_j(\beta(\varphi))
\sin^{\nu_j}(\varphi)d\varphi.
\end{equation}
where the coefficients $f_j$ indicate each population's
relative contribution to the cross--sectional density in the
ecliptic at $r_1=1$ AU.
Of course each population may also have a distinct
albedo $a_j$, in which case the $f_j$ in Eq.\ (\ref{Z_j})
should be replaced by $(a_j/a)f_j$ where $a$ is the
`effective' albedo that would be obtained if only a single dust
population was responsible for all of the zodiacal light. However
the individual albedos of asteroidal and cometary dust are not
known with any certainty so we will simply set $a_j/a=1$.

\subsubsection{source populations}
\label{sources}

The inclination distribution for the meteoritic dust complex is
related to the inclinations of the sources of that dust, namely,
colliding asteroids, active dust--producing comets,
and interstellar dust sources. The upper portion of
Fig.\ \ref{inc}
shows the inclination distribution of near--Earth, main--belt,
and Trojan asteroids. The smooth curve in this figure also shows
that the observed inclination distribution can be qualitatively
represented by the form
$g(i)\propto\sin ie^{-(i/\sigma)^2/2}$ with $\sigma=6.2^\circ$.
Note, however, that this curve seriously undercounts asteroids
with inclinations $i>20^\circ$ that are about $6\%$ of the total
asteroid population.

The inclination distributions for comets having perihelia
$q<2.5$ AU that are presumably active producers of dust 
are also shown in Fig.\ \ref{inc} where they are divided into
three dynamical classes: the Jupiter--family comets (JFCs), the
Halley--type comets (HTCs), and the Oort Cloud comets (OCCs). The
JFCs have the lowest inclinations since they likely originated
in the (relatively) low--inclination Kuiper Belt \citep{LD97}. 
Conversely, the wide--ranging OCCs have the highest inclinations
since their orbital planes have been randomized
by the galactic tide and passing stars \citep{DQT87}.
However the HTCs have intermediate inclinations as they likely
originate in a somewhat flatted inner Oort Cloud \citep{LDD01}.
The smooth curves in Fig.\ \ref{inc} also show
representative inclination distributions with $\sigma=8^\circ$
for the JFCs, $\sigma=33^\circ$ for the HTCs, and
$g(i)\propto\sin i$ for the isotropic OCCs.
Interstellar dust should also be distributed isotropically in
the inner solar system and thus have a similar inclination
distribution. Note, however, that these cometary $\sigma$'s
should be regarded as rough estimates since the
apparent orbital distributions suffer from selection
effects that tend to favor the discovery of comets in
low--perihelia, low--inclination orbits. In fact, the
underabundance of high--inclination OCCs in Figure
\ref{inc} suggests that selection effects may be
especially severe for these single--apparition comets.

Evidently, there are three classes of dust sources having
distinct inclination distributions: a lower inclination
population composed of asteroids and JFCs having a
$\sigma_{low}\simeq7^\circ$, a higher inclination
population composed of HTCs having a
$\sigma_{high}\simeq33^\circ$, and an isotropic population
of dust from OCCs and interstellar sources.
With these $\sigma$'s in hand, the normalization coefficients
$c_j$ appearing in Eq.\ (\ref{g(i)}) are obtained by
numerically integrating Eq.\ (\ref{g-h}) and requiring
$h_j(\beta=0)$ equal unity, which yields
$c_{low}=10.27$ and $c_{high}=2.190$.

The remainder of this study shall assume that the dust observed
in Fig.\ \ref{mosaic} have the same inclination distributions
as their source populations. However this need not necessarily
be true since dust grains will slowly spiral sunward due to
Poynting--Robertson (PR) drag. As a consequence of this radial
mobilization, dust grains can traverse secular resonances
with the planets where they can experience additional
inclination excitation. The degree of this excitation is
size dependent since a larger particle will drift at a slower
rate across a resonance and thus experience greater
inclination excitation. Dynamical models indicate that grains
with radii $R\gtrsim100\ \mu$m will suffer significant
inclination--pumping, {\it i.e.} $\Delta i\gtrsim 10^\circ$,
whereas grains smaller than $R\lesssim30\ \mu$m suffer only
modest pumping, {\it i.e.} $\Delta i\lesssim3^\circ$
\citep{JZ92, Detal01, GDD01}. However the
cross--sectional area of interplanetary dust
is dominated by grains having radii
$10\ \mu\mbox{m}\lesssim R\lesssim100\ \mu\mbox{m}$ which peaks
at $R\sim30 \mu$m \citep{Getal85}, so this additional source of
inclination--excitation is at most a marginally important
effect that is not modeled here. But even when a dust grain is
far from  a secular resonance, the planets secular perturbation
will still excite forced inclinations that are of order a few
degrees. These latter perturbations, which are also not
treated by our model, introduces a warp in
the zodiacal cloud whose inclination varies with the dust
semimajor axis.

\subsection{The abundance of asteroidal and cometary dust}
\label{abundances}

With the inclination distributions characterized,
Eq.\ \ref{g(i)} can be inserted into Eq.\ \ref{g-h} and the
latitude distributions $h_j(\beta)$ can be evaluated numerically
for each population. This in turn allows us to compute maps of
each population's surface brightness $Z_j(\theta,\phi)$ on a
geocentric latitude-longitude grid using Eq.\ \ref{Z_j}, and
isophotes for example populations are shown
in Fig.\ \ref{isophotes}. A
synthetic surface brightness map of the zodiacal light is
formed by selecting the power--laws $\nu_j$ for each population,
computing the $Z_j(\theta,\phi)$ maps, and then coadding the
maps with proportions $f_{low}$, $f_{hi}$ ,and $f_{iso}$.
Although the synthetic map has seven parameters, {\it i.e},
three $f_j$, three $\nu_j$, and $a\sigma_1 r_1$,
the available parameter space is quite limited since
$0\le f_j\le1$. Also recall the single power--law
evident in the ecliptic surface brightness profile, Fig.\
\ref{profiles}. The absence of a broken power--law
in the ecliptic profiles suggests either
(a) all populations have a
similar $\nu_j\simeq1.45$ radial variation, or (b) there is a
single population having $\nu_j\simeq1.45$ that contributes most
of the observed light. Note that the asteroidal dust
likely varies as $r^{-1}$ which is expected for dust
that spirals into the field of view
via PR drag\footnote{This assertion is true
when the dominant light--reflecting grains have an orbital
lifetime due to PR drag that is shorter than their collision
lifetimes, which is the case for grains having radii smaller
than about $\sim100\ \mu$m \citep{Getal85}.}. However the
cometary components should vary faster than $r^{-1}$ since
these dust grains can are produced {\it in situ} at rates that
also vary with distance $r$. Consequently we expect the three
populations to have $1\lesssim\nu_j\lesssim2$ or so.

\begin{figure}
\epsscale{1.0}
\vspace*{-24ex}\plotone{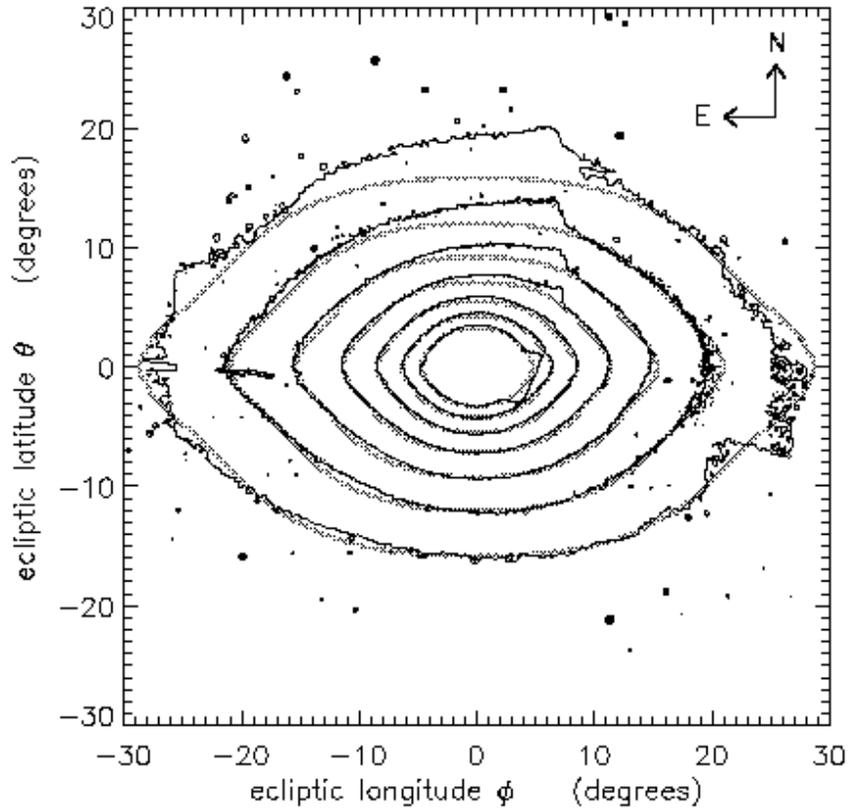}
\vspace*{-24ex}\figcaption{The narrow black curves are isophotes
of the zodiacal light map of Fig.\ \ref{mosaic}, and the thick
grey curves are isophotes for the best--fitting model having
the parameters given in Table \ref{best_fit}. The brightest
isophote corresponds to a surface brightness of
$7.45\times10^{-11}$ B$_\odot$ and successive isophotes are
dimmer by factors of 2. The outermost isophote is smoothed over
a box 5 pixels or $0.38^\circ$ wide.
The rectilinear features seen in the
outermost isophote are `seams' in the mosaic image
(see Fig.\ \ref{mosaic}); they are due to small errors in the
offsets $\delta f$ that were removed from each image (see Section
\ref{observations}). The jag in the innermost
contour as well as the linear features seen at
$\phi=-20^\circ$ and $-27^\circ$ east of the Sun are all
due to gaps in the data, and the bends in the isophotes
north--northwest of the Sun are due to scattered light in the
camera.
\label{fit}}
\end{figure}

We have scanned the $\nu_j$ parameter space between $\nu_j=1.0$
to $\nu_j=2.5$ in increments of about $\Delta\nu_j=0.5$.
It is straightforward to scan the remaining $f_j$ parameter space
for a given $\nu_j$ triplet, and in general there is only a
single set of $f_j$ parameters that agrees with the
observations in a least--squares sense. Best agreement with
the data is achieved when the $\nu_j$ and $f_j$ take the
parameters listed in Table \ref{best_fit} with the error
bars indicating the range of possible fits that are marginally
acceptable. Contours for this model are shown in
Fig.\ \ref{fit} which compares quite favorably to the observed
isophotes. However a model having all the $\nu_j=1.45$ and
$f_{low}=0.34\pm0.06$, $f_{high}=0.51\pm0.11$, 
and $f_{iso}=0.15\pm0.06$ yields isophotes that are very similar
to Fig.\ \ref{fit} with agreement that is almost as good.
This indicates that this model is particularly
sensitive to the bright, high--inclination dust population
that is distributed over a wide range of latitudes, but that the
power--law variations in the low--latitude as well dimmer
isotropic populations are less well
constrained. In general, we find acceptable solutions only when
$1.0\lesssim\nu_{low}\lesssim1.45$, $\nu_{high}\simeq1.45$, and 
$1.45\lesssim\nu_{iso}\lesssim2.0$ having approximate abundances
of $f_{low}\sim0.4$, $f_{high}\sim0.5$, and $f_{iso}\sim0.1$.

Recall that there is still a population of high inclination
asteroids having $i_{ast}\sim22^\circ$ that have not
yet been considered (see Fig.\ \ref{inc}); these
asteroids represent about $6\%$ of the total asteroid
population. Might these asteroids be a significant source of
the high latitude dust seen in Fig.\ \ref{mosaic}?
If so, then their fractional contribution to the
ecliptic surface brightness would be of order
$f_{ast}\sim6\%f_{low}\sigma_{low}/i_{ast}\sim0.01$ 
when smeared out over an annulus that is
$i_{ast}/\sigma_{low}\sim3$ times thicker than that
inhabited by the lower--$i$ asteroids. Consequently, these
high--$i$ asteroids are not likely to be a significant source of
high--latitude dust since their ecliptic contribution
is only $\sim2\%$ that of the high--inclination population's
contribution $f_{high}$.

\begin{figure}
\epsscale{1.0}
\plotone{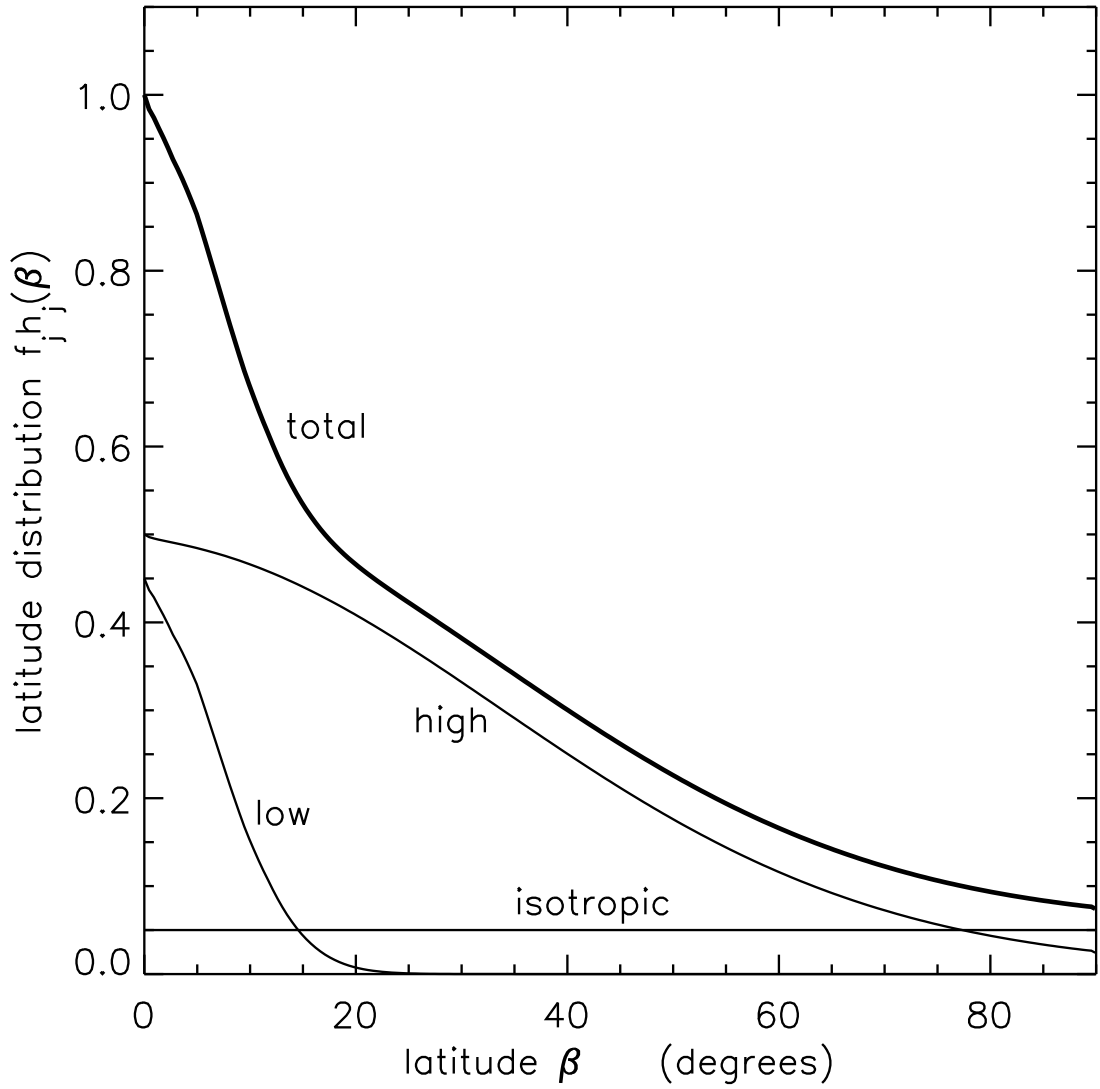}
\figcaption{The total dust latitude distribution $h(\beta)$
plotted versus heliocentric ecliptic latitude $\beta$.
Also shown are the weighted contributions by the low, high,
and isotropic populations, $f_jh_j(\beta)$.
\label{latitude}}
\end{figure}

\begin{figure}
\epsscale{1.0}
\plotone{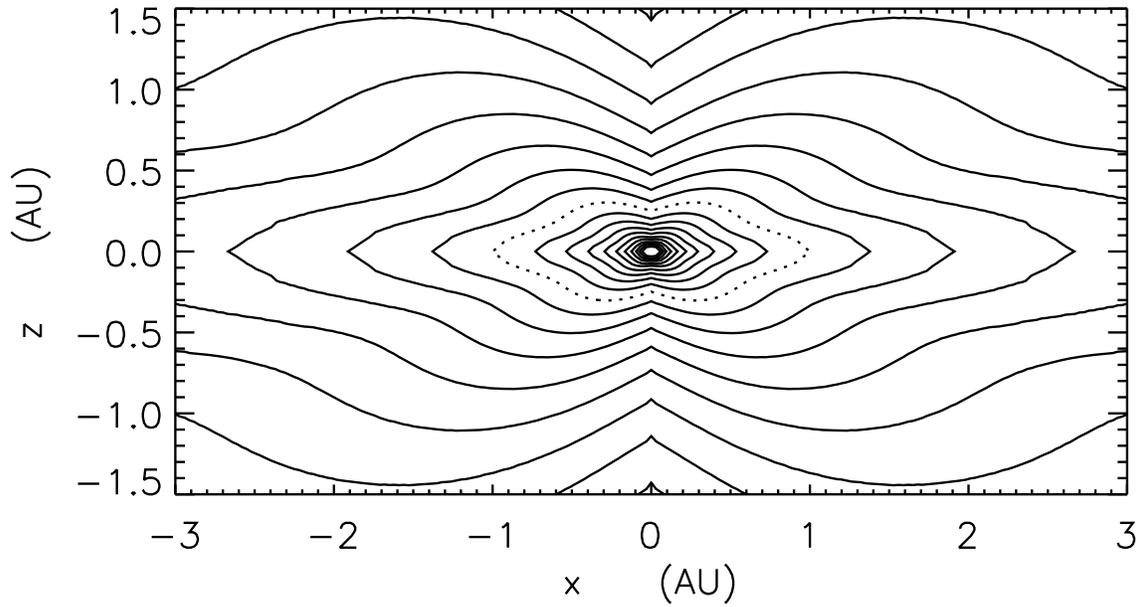}
\figcaption{Contours of the inferred density of dust
cross--section $\sigma(x,z)$ in cylindrical coordinates where
$x$ is the ecliptic distance from the Sun
and $z$ the height above the ecliptic plane. Adjacent
contours indicate a factor of 1.5 change in the dust density
and the dotted curve is where the dust density
$\sigma(x,z)=\sigma_1=4.6\times10^{-21}$ cm$^2$/cm$^3$.
\label{density}}
\end{figure}

The inferred dust latitude distribution $h(\beta)$ is
also shown in Fig.\ \ref{latitude} as well as the weighted
contributions $f_jh_j(\beta)$ by the low, high, and isotropic
dust populations. This figure shows that at ecliptic latitudes
$\beta>15^\circ$, more than $90\%$ of the cross--section is
contributed by dust that are in comet--like orbits
(e.g., HTCs and OCCs). The spatial distribution of the dust
cross section is also shown in Figure \ref{density}. Although
these contours are rather similar to that inferred by
\citet{Ketal98} from the COBE observations, they do differ in
detail due to the different assumptions built into each model.
Specifically, the Kelsall {\it et al.}\ model employs an
empirical function (namely, a modification of the
familiar fan model) to describe the dust latitude distribution
$h(\beta)$ of their `smooth cloud' (which is the principle
component of that model), whereas we allow for three distinct
distributions $h_j(\beta)$ that are instead based upon
known comet and asteroid inclination distributions. Although both
modeling efforts adopt very different treatments of the dust
latitude distributions $h(\beta)$, and these models were also
applied to data acquired at rather different wavelengths
(optical versus near and far infrared), the inferred dust
density distributions reported here and by \citet{Ketal98}
are quite similar. The density distribution given in
Fig.\ \ref{density} is also reminiscent of the familiar fan
model that assumes
$\sigma(r,\beta)=\sigma_1(r/r_1)^{-\nu}e^{-k|\sin\beta|}$.
However a parameterization of this form provides at best only a
qualitatively correct estimation of the 
density map of Fig.\ \ref{density}
when $k\simeq1.5$; it still fails to reproduce this
Figure in detail, especially at high latitudes beyond
$\beta\gtrsim45^\circ$ and at distances beyond $r\gtrsim2$ AU.

\begin{figure}
\epsscale{1.0}
\plotone{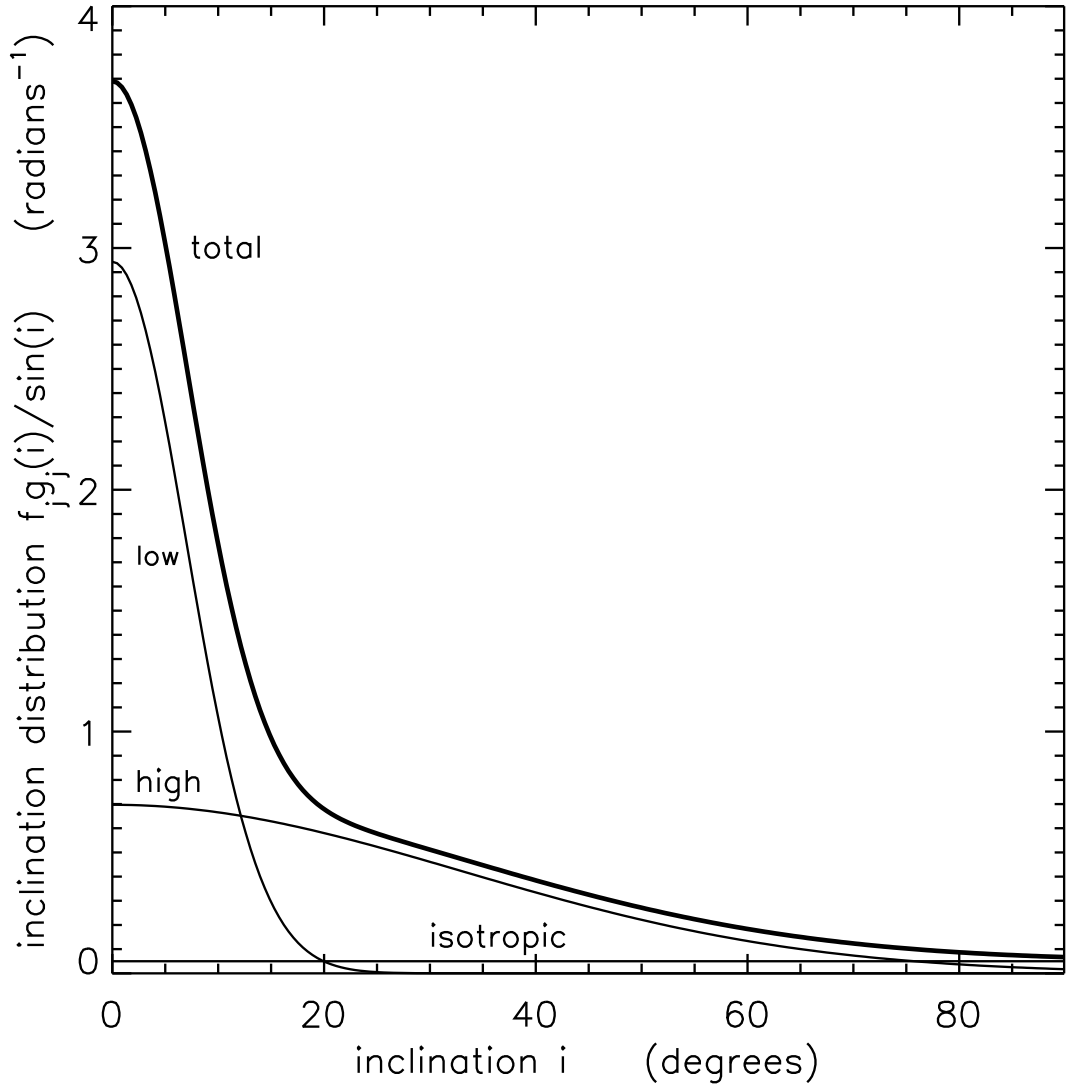}
\figcaption{The total inclination distribution $g(i)/\sin i$
plotted versus inclination $i$ as well as the weighted
contributions from the low, high, and isotropic
populations.
\label{inclination}}
\end{figure}

The inferred inclination distributions
for all of the dust populations are also shown in
Fig.\ \ref{inclination}. We also note also that
these distributions are rather similar to that reported
by \cite{D93} who inferred dust size and orbital distributions
from a wide suite of dust observations (e.g., microcraters on
lunar samples, spacecraft dust--impact experiments, as well as
other zodiacal light observations). However Divine did not
comment on the implications of this inclination distribution,
which we regard as one of the more interesting findings of this
study.

Since $\sigma_j(r,\beta)=f_j\sigma_1(r/r_1)^{-\nu_j}h_j(\beta)$
is the density of population $j$'s dust cross--section,
its total dust cross--section contained within a sphere of
radius $r$ is obtained from the volume integral
\begin{mathletters}
\begin{eqnarray}
\Sigma_j(r) &=& \int\sigma(r,\beta)dV \\
\label{Sigma0}
&=& \frac{f_j\gamma_j}{3-\nu_j}
\left(\frac{r}{r_1}\right)^{3-\nu_j}4\pi\sigma_1r_1^3
\end{eqnarray}
\end{mathletters}
where the latitude integration is
$\gamma_j\equiv\int^{\pi/2}_0h_j(\beta)\cos\beta d\beta$,
which numerically integrates to $\gamma_{low}=0.143$,
$\gamma_{high}=0.619$, and $\gamma_{iso}=1.00$.
Adopting the best--fitting $\nu_j$ and $f_j$ parameters given
in Table \ref{best_fit}, each population's total dust
cross--section interior to $r_1=1$ AU is
$\Sigma_j(r_1)=\{0.032, 0.200, 0.050\}\times4\pi\sigma_1r_1^3$,
respectively, for the low, high, and isotropic populations. 
Although the low--inclination dust from asteroids and
JFCs contributes $f_{low}=45\%$ of the dust cross--sectional 
density in the ecliptic, at least $89\%$ of the dust interior to
a $r_1=1$ AU sphere is contributed by sources in comet--like
orbits (e.g., HTCs and OCCs plus an unknown fraction from JFCs).

Using data from spacecraft dust collection experiments as well as
studies of lunar microcraters, \citet{Getal85} estimate
the spatial density of the
ecliptic dust cross section at 1 AU
to be $\sigma_1\sim4.6\times10^{-21}$ cm$^2$/cm$^3$.
Since Table \ref{best_fit} reports
$a\sigma_1 r_1=(7.4\pm1.8)\times10^{-9}$, this implies that the
dust have an effective albedo of $a\simeq0.1$. Note, however,
that these estimates for $\sigma_1$ and
thus $a$ are probably uncertain by a factor of $\sim2$ since
these dust collection experiments largely measure dust fluxes
versus particle energy, and that their conversion to a
dust cross--section requires assumptions about the
dust velocities and their bulk densities.

Summing Eq.\ (\ref{Sigma0}) over all populations yields
the total dust cross--section contained within a sphere of radius
$r$:
\begin{equation}
\label{Sigma1}
\Sigma(r)=2.0\times10^{10}\left[
0.032\left(\frac{r}{r_1}\right)^2+
0.200\left(\frac{r}{r_1}\right)^{1.55}+
0.050\left(\frac{r}{r_1}\right)\right]\mbox{ km}^2
\end{equation}
where each term gives the contribution by the low $i$, high $i$,
and isotropic populations, respectively. Note that this
expression only applies interior to the dust--producing
portion of the asteroid belt, {\it i.e.}\ interior to
$r\simeq3.3$ AU
\citep{Hetal74}. If we consider a sphere of radius $r_2=2$ AU
enclosing the orbits of the terrestrial planets, then
$\Sigma(r_2)=1.6\times10^{10}$ km$^2$, which 
is about 50 times the total cross--section of the terrestrial
planets. This estimate illustrates one of the main difficulties
challenging efforts to detect extra--solar
planets via direct imaging or interferometry at optical
wavelengths: if terrestrial
extra--solar planets are also embedded in solar system--like
dust, then one will need to resolve planetary systems to fairly
small spatial scales in order to discriminate the starlight
reflected by planets from that reflected by dust.

The Pioneer 10 spacecraft detected asteroidal dust out to a
heliocentric distance of $r_3\simeq3.3$ AU \citep{Hetal74},
so Eq.\ (\ref{Sigma0}) indicates that the low--inclination dust
component has a total surface area
$\Sigma_{low}(r_3)=6.8\times10^9$ km$^2$.
Note that observations of the
IRAS dust bands serve as a comforting reality check on our
findings since the total surface area associated with the three
most prominent asteroidal dust bands is
$\Sigma_{band}=4.7\times10^9$ km$^2$ \citep{GDD01};
the remaining dust must then
be due to other minor asteroid families,
non--family asteroids, and JFC comets. Also, 
if the light--reflecting dust seen in Fig.\ \ref{mosaic}
can be attributed to grains having a
characteristic radius $R_c$, then the total number of grains
interior to distance $r$ is
$N_j(r)\sim\Sigma_j(r)/\pi R_c^2$
and their enclosed mass
$M_j(r)\sim4\rho R_c\Sigma_j(r)/3$ is
\begin{equation}
\label{Mj}
M_j(r)\sim
\frac{16\pi R_c\sigma_1\rho r_1^3f_j\gamma_j}{3(3-\nu_j)}
\left(\frac{r}{r_1}\right)^{3-\nu_j}
\end{equation}
where $\rho$ is the grains' bulk density. In this case the total
mass of the  light--reflecting component of asteroidal dust is
at most of order
\begin{equation}
M_{low}(r_3)\sim2.3\times10^{18}
\left(\frac{\rho}{\mbox{2.5 gm/cm$^3$}}\right)
\left(\frac{R_c}{100\ \mu\mbox{m}}\right)\mbox{ gm}.
\end{equation}
The interplanetary dust mass distribution peaks at
$R_c\sim100\ \mu$m \citep{Getal85}, so the above dust
mass--limit is equivalent to an asteroid that is about
12 km across. Of course this limit is valid only if our model,
which is based on observations of dust orbiting at $r\lesssim0.6$
AU, can be reliably extrapolated out to the asteroid belt.
Nonetheless, the similarity between our dust model
and that inferred from the COBE observations, which are sensitive
to dust in the $1\lesssim r\lesssim3$ AU interval
\citep{Ketal98}, indicate that our
extrapolation is indeed valid. We also
note that the mass limit obtained here about 3.5 times
the mass of the asteroidal dust bands
detected by COBE \citep{Retal97}, but keep in mind that our
limit is also contaminated by dust from JFCs.

The isotropic cloud of dust is also quite interesting,
and the following discussion assesses the relative dust
contribution from Oort Cloud comets versus interstellar sources.
The dust seen in Fig.\ \ref{mosaic} having a typical elongation
of $\epsilon\sim15^\circ$ orbit at a heliocentric distance of
$r\sim r_1\sin\epsilon\sim0.3$ AU, and this dust has a
cross--sectional density of $\sigma_{iso}(0.3\mbox{ AU})=f_{iso}
\sigma_1(0.3\mbox{ AU}/r_1)^{-\nu_{iso}}
\sim3\times10^{-21}$ cm$^2$/cm$^3$ in the ecliptic.
The interstellar fraction
is inferred from the flux of interstellar dust measured by
impact detectors onboard the Galileo and Ulysses spacecraft.
These detectors measured an interstellar dust flux of
$f_\star=1.5\times10^{-8}$ grains/cm$^{2}$/sec
[after correcting a typo in \cite{Getal97}]. These interstellar
grains have a mean mass of $m_\star\sim3\times10^{-13}$ gm, so
their characteristic radius is $R_\star\sim0.3\ \mu$m.
It will be assumed here that the interstellar dust flux is
roughly constant throughout the solar
system since radiation pressure roughly balances solar gravity
for grains of this size. Since interstellar matter approaches the
solar system with a
velocity--at--infinity of $v_\infty=25$ km/sec
\citep{Frisch00}, the number density of interstellar dust is
$n_\star\sim f_\star/v_\infty\sim6\times10^{-15}$ cm$^{-3}$
and their cross--sectional density is
$\sigma_\star\sim\pi R_\star^2n_\star\sim2\times10^{-23}$
cm$^2$/cm$^3$. This is only about $1\%$ of the observed
cross--sectional density $\sigma_{iso}$, which indicates
that the isotropic portion of the dust seen in Fig.\ \ref{mosaic}
comes predominantly from Oort Cloud comets.

Oort Cloud comets have semimajor axes
$a\sim10^4$ AU, and those comets passing sufficiently close to
the Sun will sublimate gas and dust that gets injected into
orbits similar to their parent comets. If we
naively extrapolate these wide--ranging dust grains out to Oort
Cloud distances using the inferred $\nu_{iso}=2$ radial
power--law, the total Oort Cloud dust mass is
\begin{equation}
\label{Miso}
M_{iso}\sim1\times10^{19}
\left(\frac{\rho}{\mbox{1 gm/cm$^3$}}\right)
\left(\frac{R_c}{1\ \mu\mbox{m}}\right)
\left(\frac{a}{10^4\mbox{ AU}}\right)
\mbox{ gm},
\end{equation}
which has a equivalent to a $\sim30$ km comet. However this mass
is easily uncertain by orders of magnitude due to uncertainties
in the size and bulk density of the dust as well as the radius of
the Oort Cloud. Ultimately these distant dust
grains will be stripped from the solar system as they are swept
up by the interstellar gas and dust that flows through
the solar system. Thus it is conceivable that the Sun also
has a vast but tenuous tail of Oort Cloud dust. If so, this dust
tail would be oriented in the downstream direction of the local
interstellar flow which has a heliocentric
ecliptic longitude, latitude
of $(74.7^\circ, -4.6^\circ)$ \citep{Frisch00} or an
equatorial right ascension, declination of
$(73.9^\circ, +18.0^\circ)$.

\section{Summary and Conclusions}
\label{summary}

Using the Moon to occult the Sun, the Clementine spacecraft used
its navigation cameras to map the inner zodiacal light at optical
wavelengths over elongations of
$3\lesssim\epsilon\lesssim30^\circ$ from the Sun. Since 
the zodiacal light is sunlight that is reflected by
interplanetary dust, this map provides a measure of the dust
grains' radial and vertical variations spanning heliocentric
distances of $0.05\lesssim r\lesssim0.6$ AU, {\it i.e.}, from
about 10 solar radii to just interior to Venus' orbit. The
integrated zodiacal light seen over the
$60^\circ\times60^\circ$ field of view has a visible magnitude
$m_V=-8.5$, indicating that the meteoritic complex is
one of the brightest members of the planetary system,
second only to the full Moon.

The averaged ecliptic surface brightness of the zodiacal light
falls off as $Z(\epsilon)\propto\epsilon^{-2.45\pm0.05}$ which
suggests that the dust cross--sectional density nominally varies
as $\sigma(r)\propto r^{-1.45\pm0.05}$ (but see below).
This surface brightness also indicates that the dust obey
$a\sigma_1r_1=(7.4\pm1.8)\times10^{-9}$. Assuming that the
dust have an ecliptic cross--sectional density of
$\sigma_1=4.6\times10^{-21}$ cm$^2$/cm$^3$ at $r_1=1$ AU
\citep{Getal85}, this implies that the dust have
an effective albedo of $a=0.1$
that is perhaps uncertain by a factor of 2. Asymmetries of
$\sim10\%$ in the zodiacal light's surface brightness 
are evident at elongations $\epsilon\sim15^\circ$ in
directions east--west as well as north--south of the Sun,
and these asymmetries may be due to the giant planets' secular
gravitational
perturbations. However simple digital filtering of the data
({\it i.e.}, unsharp masking) does not reveal any other subtle
features such as dust bands associated with
asteroid families or dust trails associated with
individual comets.

In order to assess the relative contributions to the
interplanetary dust complex by asteroids and comets,
we have modeled the zodiacal cloud as being due to three dust
populations having distinct inclination distributions.
One dust population is assumed to have low inclinations
that are distributed as a gaussian
with a standard deviation $\sigma_{low}\simeq7^\circ$ that is
characteristic of both asteroids as well as Jupiter--Family
comets (JFCs). A higher--inclination population corresponding to
the Halley--type comets (HTCs) is assumed to have
a $\sigma_{high}\simeq33^\circ$, and the third population is an
isotropic cloud of dust from Oort Cloud comets. This simple
model is applied to the observations and very good agreement is
achieved for the parameters listed in Table \ref{best_fit}
(see Fig.\ \ref{fit}). It should be noted that the
best--fitting model has a radial power--law
$\nu_{low}=1.0$ for the asteroidal + JFC population, which is
consistent with dust delivery via Poynting--Robertson (PR) drag
yet shallower than the nominal $\nu=1.45$
power--law quoted above. Also, the dust from HTC appears to
follow a $\nu_{high}=1.45$ power--law while the Oort Cloud dust
varies as $\nu_{iso}=2.0$. Interstellar dust also contributes to
this isotropic cloud, but only at the $\sim1\%$ level.
Yet despite this mixture of power--laws,
the resulting surface brightness
profile still varies close to the observed
$Z(\epsilon)\propto\epsilon^{-2.45}$. Note, however,
that acceptable agreement with the data is also achieved when all
populations have $\nu_j=1.45$; see Section \ref{abundances} for
the allowed range of model parameters.

The best--fitting model indicates that about $f_{low}=45\%$
of the dust cross--section in the ecliptic at $r_1=1$ AU comes
from asteroids and JFCs. But when a 1 AU--radius sphere is
considered, at least $89\%$ of the integrated
dust cross section comes from
sources in comet--like orbits. However it should be noted that
these findings are inferred from a `static' model of stationary
dust grains. This rather simple approach is applicable provided
dust grains (and in particular, asteroidal dust) do not
experience substantial--inclination pumping
as they evolve sunwards due to PR drag and cross orbital
resonances with the planets.
Although this appears to be the case for grains smaller than
$\sim100\ \mu$m that are the dominant source
of reflected sunlight, inclination pumping is certainly of
greater importance for the larger dust grains that can drift
across resonances at
slower rates. In this instance, `dynamic' models that include
radiation and gravitational forces are preferred as they can
faithfully follow a dust grain's orbital evolution from source
to sink, and these more sophisticated models have been used to
extract dust properties from the infrared observations of the
outer zodiacal light acquired by IRAS
(cf.\ \cite{Detal01,GDD01}). Of
course these dynamic dust models are
also applicable to the optical observations of the inner
zodiacal light examined here, and we suspect that any effort to
simultaneously fit a dynamic dust model to the full suite optical
and infrared IRAS, COBE, and Clementine observations should yield
an even more tightly constrained picture of the interplanetary
dust complex spanning a very wide range of heliocentric distances
$0.05\lesssim r\lesssim3.3$ AU. And in order to facilitate any
such effort, the Clementine map of the inner zodiacal light
is available from the authors by request.

The inferred cross--section of dust orbiting interior to 2 AU is
about 50 times that of the terrestrial planets. This
suggests that any effort to directly detect terrestrial
extrasolar planets at optical wavelengths is faced with the
daunting task of distinguishing the faint starlight reflected
by such planets from the far brighter signal anticipated
from any exozodiacal dust. When these results are
extrapolated out into the asteroid belt, the total mass of the
light--reflecting asteroidal dust component is at most
$\sim2.3\times10^{18}$ gm. Note that this mass--limit is also
contaminated by dust from JFCs, and it corresponds to an asteroid
that is at most $\sim12$ km across.

When these results are extrapolated out to Oort Cloud
distances of $a\sim10^4$ AU, the inferred mass of Oort Cloud
dust is $\sim10^{19}$ gm (but uncertain by orders of magnitude),
which is equivalent to a 30 km comet.
This dust is ultimately stripped from the Sun by the
interstellar gas and dust that flows around and through the
solar system. This then suggests that the Sun, and perhaps
also other stars having cometary Oort Clouds, each
have vast but tenuous stellar dust tails that are oriented in the
downstream direction of the local interstellar flows.

\newpage
\appendix
\section{Appendix \ref{appendixA}}
\label{appendixA}

To relate a pixel's $(x,y)$ coordinates to the right ascension
and declination $(\alpha,\delta)$ it subtends on the sky,
place a Cartesian coordinate system with its origin at the
center of the lens with the ${\bf \hat{x}}$, ${\bf \hat{y}}$
axes parallel to the CCD's rows and columns
(see Fig.\ \ref{optics}).
The ${\bf \hat{z}}$ axis is the camera's optical axis
which intercepts the CCD at the pixel having coordinates
$(x_o,y_o)=(191,286)$. Let the vector
${\bf r}_\star$ point to a star having angular coordinates
$(\theta,\phi)$; these will be called the lens coordinates.
In cartesian $(x,y,z)$ components, the star's lens coordinates 
are ${\bf r}_\star=(R\sin\phi\cos\theta, R\sin\theta,
R\cos\phi\cos\theta)$, and the length of this vector is chosen
to be the lens radius $R$. An image of this star also forms at
the back side of the lens at $-{\bf r}_\star$. 
Fig.\ \ref{optics} shows that the fiber optic
pipes this starlight to a spot on the CCD that lies a distance
$X$ away from the optical axis in the ${\bf \hat{x}}$ direction
and $Y$ away in the $-{\bf \hat{y}}$ direction. Note
that the orientation of the $X$ and $Y$ axes are chosen to point
in the customary manner such that
when ${\bf \hat{y}}$ points to equatorial north, the $+Y$
direction is north and the $+X$ direction is west.
If $l$ is the physical size of a pixel,
then $X=l(x-x_o)=-{\bf r}_\star\cdot{\bf \hat{x}}$,
$Y=l(y-y_o)={\bf r}_\star\cdot{\bf \hat{y}}$, and
\begin{mathletters}
\label{xy}
\begin{eqnarray}
x-x_o &=& -\sin\phi\cos\theta/p\\
y-y_o &=& \sin\theta/p
\end{eqnarray}
\end{mathletters}
relates the star's $(x,y)$ coordinates on the CCD to its
lens coordinates $(\theta,\phi)$. The
plate--scale $p=\ell/R$ is simply the ratio of the pixel width
to the lens radius, and is also the angle subtended by the pixel
at the optical axis. 

Now relate the lens coordinates $(\theta,\phi)$ to 
equatorial coordinates $(\alpha,\delta)$. Suppose
the optical axis ${\bf \hat{z}}$ points to right ascension
$\alpha_o$ and declination $\delta_o$, and the
${\bf \hat{y}}$ axis differs from equatorial north
by a rotation about the ${\bf \hat{z}}$ axis by angle $\tau$.
This angle is the position angle of equatorial
north, and on the CCD it is measured from the $Y$ axis
towards the east. In this coordinate system
the star's cartesian components are
${\bf r}_\star(\alpha,\delta)=(R\sin\alpha\cos\delta,
R\sin\delta, R\cos\alpha\cos\delta)$ when expressed in terms of
its equatorial coordinates.
It is straightforward to show that this star's position
vector ${\bf r}_\star(\alpha,\delta)$ in equatorial coordinates
can be obtained from the its position vector in lens
coordinates ${\bf r}_\star(\phi,\theta)$ after performing
the following rotations upon the lens coordinate system:
\begin{equation}
\label{rotations}
{\bf r}_\star(\alpha,\delta)=
{\cal R}_y(-\alpha_o){\cal R}_x(\delta_o){\cal R}_z(-\tau)
{\bf r}_\star(\theta,\phi)
\end{equation}
where ${\cal R}_i(\omega)$ is the matrix that rotates a
right--handed coordinate system about axis $i$ by angle
$\omega$. Performing the rotations and equating the $(x,y,z)$
components in Eq.\ (\ref{rotations}) yields the relationship
between a pixel's lens coordinates $(\theta,\phi)$ and its
equatorial coordinates $(\alpha,\delta)$:
\begin{mathletters}
\label{a_o,d_o}
\begin{eqnarray}
\sin\delta &=& g\cos\delta_o + h\sin\delta_o\\
\tan\alpha &=& \frac{k\sin\alpha_o+f\cos\alpha_o}
{k\cos\alpha_o-f\sin\alpha_o}
\end{eqnarray}
\end{mathletters}
where $f,g,h$, and $k$ are shorthand for
\begin{mathletters}
\begin{eqnarray}
f &=& \cos\tau\sin\phi\cos\theta-\sin\tau\sin\theta \\
g &=& \sin\tau\sin\phi\cos\theta+\cos\tau\sin\theta \\
h &=& \cos\phi\cos\theta \\
k &=& h\cos\delta_o-g\sin\delta_o.
\end{eqnarray}
\end{mathletters}

Next, get the star's geocentric ecliptic longitude and latitude
$(\Lambda,\Theta)$ where\linebreak
${\bf r}_\star(\Lambda,\Theta)=(R\sin\Lambda\cos\Theta,
R\sin\Theta, R\cos\Lambda\cos\Theta)$. These are
obtained by rotating the ecliptic coordinates about the $z$ axis
by the Earth's obliquity $\sigma=23.4393^\circ$, {\it i.e.},
${\bf r}_\star(\Lambda,\Theta)=
{\cal R}_z(\sigma){\bf r}_\star(\alpha,\delta)$.
Equating cartesian components yields
\begin{mathletters}
\label{lat-long}
\begin{eqnarray}
\sin\Theta &=&
\cos\sigma\sin\delta-\sin\sigma\sin\alpha\cos\delta\\
\tan\Lambda &=&
\label{longitude}
\frac{\cos\sigma\sin\alpha\cos\delta+\sin\sigma\sin\delta}
{\cos\alpha\cos\delta}.
\end{eqnarray}
\end{mathletters}
This gives the star's longitude measured from the vernal
equinox. To get the star's longitude relative to the Sun, use
Eq.\ (\ref{longitude}) to compute the Sun's longitude
$\Lambda_\odot$ and form the difference $\Lambda-\Lambda_\odot.$

The observed angular separation $\Phi_{01}$
between two known stars
can be used to determine the detector's plate--scale $p$. If the
stars have lens coordinates $(\theta_0,\phi_0)$ and
$(\theta_1,\phi_1)$, which depend on $p$, and equatorial
coordinates $(\alpha_0,\delta_0)$ and $(\alpha_1,\delta_1)$,
then spherical geometry gives their angular separation:
\begin{mathletters}
\label{Phi_01}
\begin{eqnarray}
\cos\Phi_{01} &=& \cos\theta_0\cos\theta_1\cos(\phi_0-\phi_1)
+\sin\theta_0\sin\theta_1 \\
             &=& \cos\delta_0\cos\delta_1\cos(\alpha_0-\alpha_1)
+\sin\delta_0\sin\delta_1.
\end{eqnarray}
\end{mathletters}
This equation can be solved for $p$ when coupled with
Eqs.\ (\ref{xy}). Another useful quantity is the angular
separation between star 0 and the optical axis at
$(\theta_1,\phi_1)=(0,0)$,
\begin{equation}
\label{Phi_0}
\cos\Phi_0=\cos\theta_0\cos\phi_0.
\end{equation}
Using Fig.\ \ref{optics}, it can also be shown that an off--axis
pixel subtends a larger solid angle $\Omega(\Phi_0)$ than the
on--axis pixel:
\begin{equation}
\label{solid-angle}
\frac{\Omega}{\Omega}_0=\frac{1}{\cos\Phi_0}
\end{equation}
where $\Omega_0=p^2$ is the solid angle of the pixel at the
optical axis. For example, a pixel in the CCD corner at
$\Phi_0=25^\circ$ sees a solid angle that is
$10\%$ larger than the on--axis pixel.

\section{Appendix \ref{appendixB}}
\label{appendixB}

\begin{figure}
\epsscale{1.0}
\vspace*{-3ex}\plotone{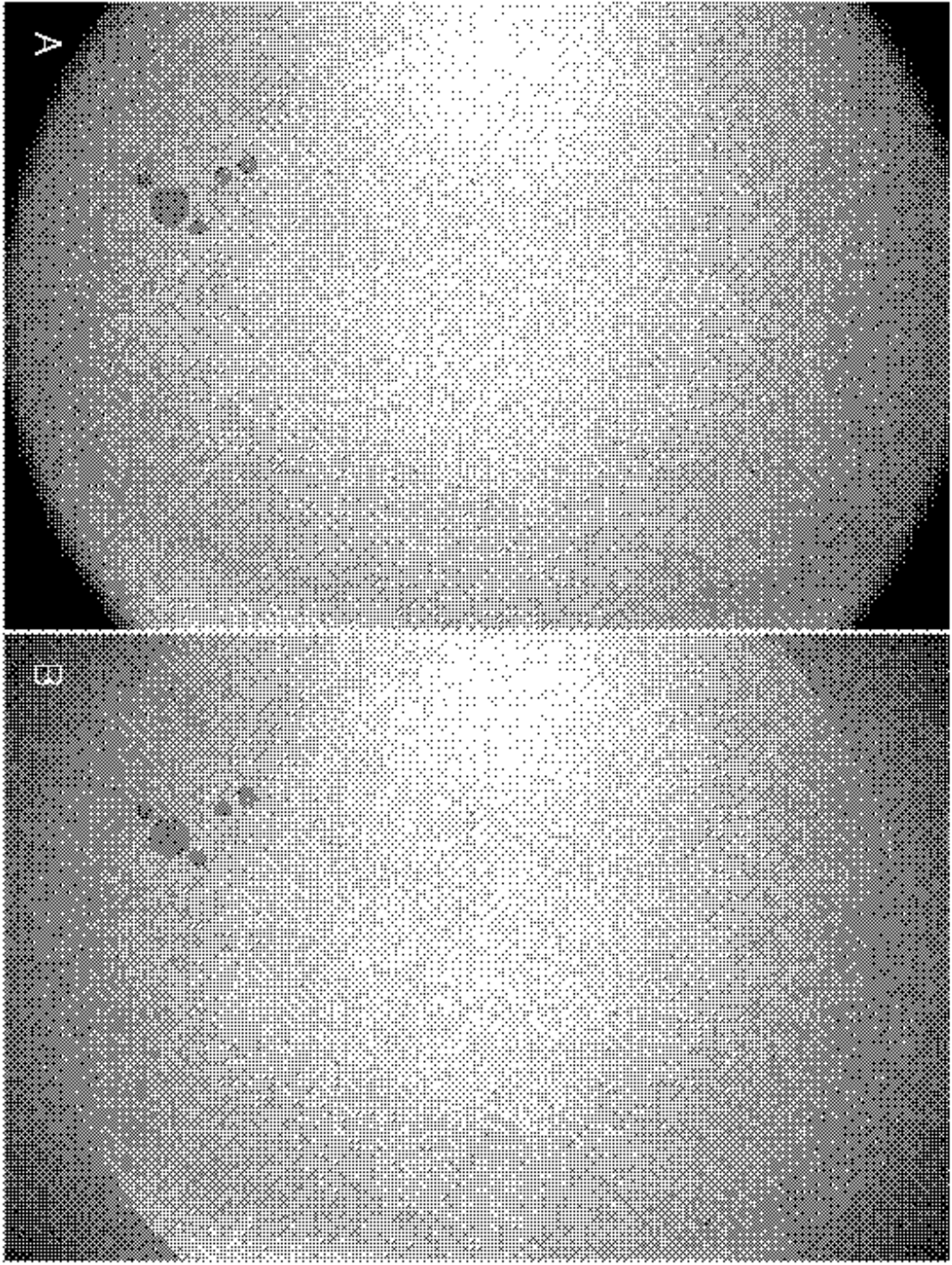}
\vspace*{-7ex}\figcaption{{\bf A}. The vignetted flatfield.
{\bf B}. The reconstructed flatfield.
\label{flat}}
\end{figure}

Flatfield data for camera B were acquired prior to launch in a
laboratory at Research Support Instruments, which is the
subcontractor that integrated the star trackers into the
Clementine payload. These images were acquired with star tracker
B exposed to an integrating sphere, which is a spherical
light source having a uniform radiance. The averaged
image acquired in the lab is shown in Fig.\ \ref{flat}A.
The following describes several problems in these data,
as well as the procedures used to construct
the star tracker flatfield from these data.

({\it i}.) The integrating sphere was too bright to produce
unsaturated images at the star tracker's shortest possible
exposure time. To get unsaturated images, a Wrattan neutral
density ND2 filter was placed in front of the camera. It is
thus conceivable that if there are any spatial variations
$\delta T(x,y)$ in this filter's transmission coefficient
$T(x,y)$, the resulting flatfielded data would then
exhibit fractional errors $\delta T/T$. However
we have ruled out this possibility by
examining distinct yet overlapping zodiacal light fields 
acquired during different orbits with different camera
orientations. In every case, the photometry in each overlapping,
flatfielded image was self--consistent. This
indicates that any errors $\delta T(x,y)$ that may have been
introduced into the flatfield by the use of the neutral density
filter are negligible.

({\it ii}.) The dark current was not measured when the flatfield
data were acquired. This could have been quite problematic since
the the subtraction of the dark current has a dramatic affect
upon the brightness of the flatfield center relative to its
edges (see Fig.\ \ref{flat}). However a good flatfield is still
recoverable due to the fortunate presence of the five round
blemishes seen in Fig.\ \ref{flat}. These blemishes
correspond to a $\sim10\%$ decrease in the camera's
sensitivity, and their cause is unknown.
These blemishes are present in all of the raw zodiacal light
images acquired by star tracker B ({\it c.f}.\ the
lower right corner of Fig.\ \ref{raw}A). However the
relative depth of the blemish increases with increasing dark
current $d$, so it is straightforward to find the
appropriate value for $d$ that yields a flatfield that
removes the blemishes from all of the zodiacal light images.
Note that the flatfielded image in Fig.\ \ref{raw}B
does not exhibit these blemishes.

({\it iii}.) Vignetting by the mount that
secured the neutral density
filter to the camera also darkened the outer edges of the
flatfield data (see Fig.\ \ref{flat}A). Interior to the
vignetted region, the flatfield's surface brightness
empirically varies as a simple polynomial
\begin{equation}
\label{flatSB}
\mbox{flat}(x,y)=1-1.95\Phi_0^2(x,y)
\end{equation}
that is a function of the angular distance $\Phi_0$ of pixel
$(x,y)$ from the the optical axis (see Eq.\ \ref{Phi_0}).
In order to reconstruct the flatfield at the edges, Eq.\
(\ref{flatSB}) is extrapolated into the vignetted region
at $\Phi_0>20^\circ$,
which results in the flatfield shown in Fig.\ \ref{flat}B.
Gaussian noise is also added to the extrapolated
region in amounts comparable to that seen just interior.

The pixel--to--pixel variations seen over short spatial
scales in Fig.\ \ref{flat} is simply photon--counting
noise---they do not represent real changes in the camera's
sensitivity. This becomes evident when flatfielding these
images. Ordinarily, flatfielding an image makes it look
`cleaner' since the pixel--to--pixel variations
in the CCD's sensitivity are removed from the image.
However the application of the
flatfield seen in Fig.\ \ref{flat}B actually makes the zodiacal
light images look a little bit noisier.
It may be concluded that the
short--wavelength variations seen in the flatfield are simply
noise, and that we are justified in adding gaussian noise to the
extrapolated parts of the flatfield. (Failure to add this noise
creates a lower--noise zone at the edge of a flatfielded image
having a very artificial appearance.)
The standard deviation of this noise near the optical axis
is about $2\%$ of the flatfield's surface brightness, and it
increases to about $7\%$ at the edge. 

Figure \ref{flat} and Eq.\ (\ref{flatSB})
show that the surface brightness of the
flatfield decreases with distance from the optical axis by
as much as $40\%$ at the CCD corners where $\Phi_0=25^\circ$.
This seemingly runs contrary to expectations
since Appendix \ref{appendixA} shows that pixels further from
the optical axis subtend a solid angle that is larger by
a factor of $1/\cos(\Phi_0)$ (Eq.\ \ref{solid-angle}), and
perhaps should be brighter. However
this effect is more than offset by transmission inefficiencies
in the fiber--optic and also vignetting by the lens housing,
both of which get more severe for light entering the lens at
larger angles \citep{Letal91}. Consequently, 
dividing a raw image by this flatfield simultaneously
corrects for two effects: it compensates for the camera's reduced
detection efficiency at higher $\Phi_0$ as well as
for the pixels' larger solid angle. 

Because the reconstructed flatfield of Fig.\ \ref{flat}B
compensates for a pixel's larger solid at higher $\Phi_0$, it
should only be used when doing photometry on distributed sources
such as the zodiacal light. But if photometry on point--sources
is desired, a ``point--source'' flatfield, which is just
Fig. \ref{flat}B $\times\cos(\Phi_0)$, must instead be used to
flatfield the images.

\begin{figure}
\epsscale{1.0}
\plotone{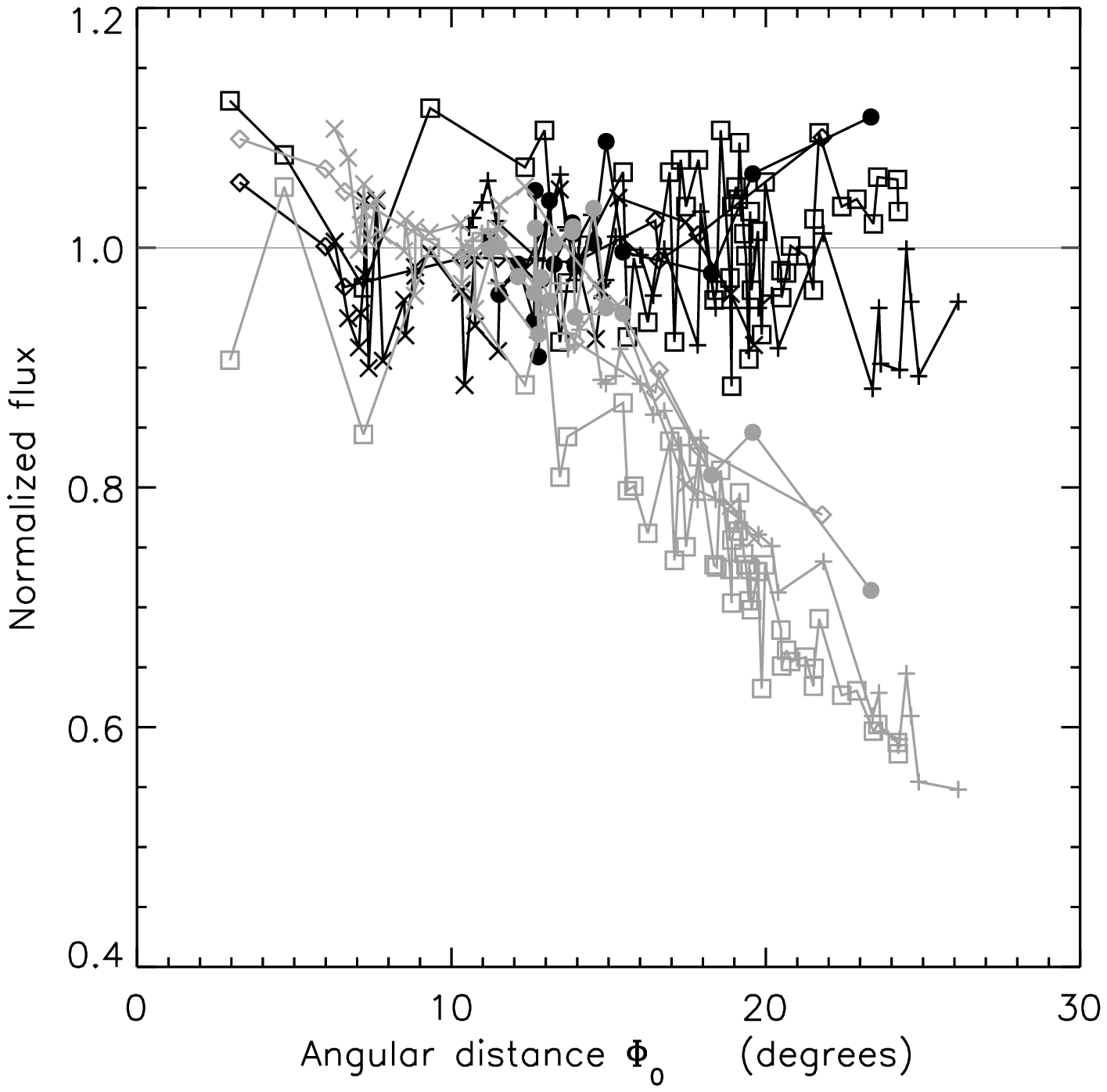}
\figcaption{The intensities of five stars
(indicated by different plotting symbols) seen in different
images acquired during orbit 66 are plotted versus each star's
angular distance from the optical axis $\Phi_0$.
All intensities are normalized to unity near $\Phi_0=0$.
The grey curves are extracted from images that are not
flatfielded, and the dark curves are from images
obtained with the reconstructed point--source flatfield.
\label{flat_test}}
\end{figure}

The observed intensity of several stars imaged
during orbit 66 are used to test this reconstructed flatfield.
The spacecraft was maneuvering while this image sequence was
acquired, so stars seen in different images wander
across the field. Figure \ref{flat_test} plots the normalized
intensity of five bright stars seen in different images versus
their angular distance $\Phi_0$ from the optical axis.
The grey curves are their intensities prior to flatfielding,
which shows how the camera's detection efficiency drops with
$\Phi_0$. The black curves shows the intensities of these stars
when the point--source flatfield is used. This test demonstrates
that when the reconstructed flatfield is used, the stars'
intensities are constant when imaged at different
position on the CCD, as they should.

\acknowledgements

\begin{figure}[t]
\epsscale{1.0}
\vspace*{-0.75in}\plotone{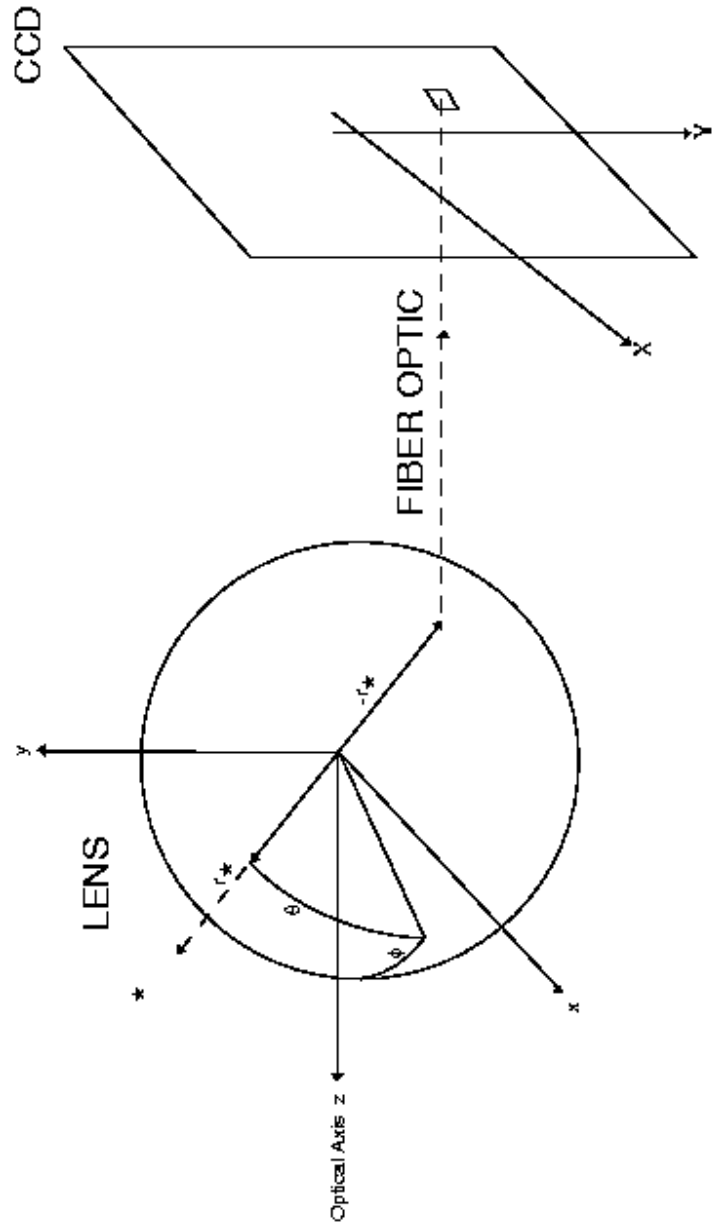}\vspace*{-0.75in}
\caption{A schematic of the star tracker camera which has three
principle components: a spherical lens and a fiber optic that
pipes light from the backside of the lens to the CCD detector.}
\label{optics}
\end{figure}

This paper is dedicated to the memory of Herb Zook, without
whom none of this would have been possible. The authors also
thank Paul Spudis who, as the Clementine deputy scientist,
allowed the spacecraft's navigation cameras to be pointed
sunwards. The authors also thank
Julie Moses for helpful discussions on interactions
between interplanetary dust and the interstellar medium and
Ronna Hurd for composing Fig.\ \ref{optics}. The authors also
thank William Reach and an anonymous reviewer for comments
that led to several improvements in this paper.
This paper is contribution \#1117 from the Lunar and Planetary
Institute which is operated by the Universities Space Research
Association under NASA contract NASW--4574.


\newpage
\begin{deluxetable}{cccc}
\renewcommand{\arraystretch}{0.7}
\tablenum{I}
\tablewidth{0pt}
\tablecaption{Zodiacal Light Observations\label{orbits}}
\tablehead{
  \colhead{Orbit} & \colhead{Date} & \colhead{Ecliptic} &
    \colhead{Total}\\
  \colhead{number} & \colhead{(1994)} & \colhead{longitude} & 
    \colhead{exp.\ time}\\
  \colhead{} & \colhead{} & \colhead{(degrees)} &
    \colhead{(seconds)} 
}
\startdata
66  & March  5.9 & 345.1 &  2.6 \\
110 & March 15.0 & 354.2 &  0.3 \\
110 & March 15.0 & 354.3 &  0.6 \\
164 & March 26.4 &   5.5 &  8.7 \\
193 & April  1.5 &  11.5 & 11.6 \\
206 & April  4.0 &  14.0 & 11.4 \\
253 & April 13.9 &  23.7 & 13.9
\enddata
\end{deluxetable}

\newpage
\begin{deluxetable}{lcc}
\renewcommand{\arraystretch}{0.7}
\tablenum{II}
\tablewidth{0pt}
\tablecaption{Parameters for the Best Fit\label{best_fit}}
\tablehead{
  \colhead{$j$} & \colhead{$\nu_j$} & \colhead{$f_j$}
}
\startdata
low  & 1.00 & $0.45\pm0.13$ \\
high & 1.45 & $0.50\pm0.02$ \\
iso  & 2.00 & $0.05\pm0.02$ \\
\hline
     & $a\sigma_1r_1=(7.4\pm1.8)\times10^{-9}$
\enddata
\end{deluxetable}

\newpage

\begin{figure}
\epsscale{0.7}
\plotone{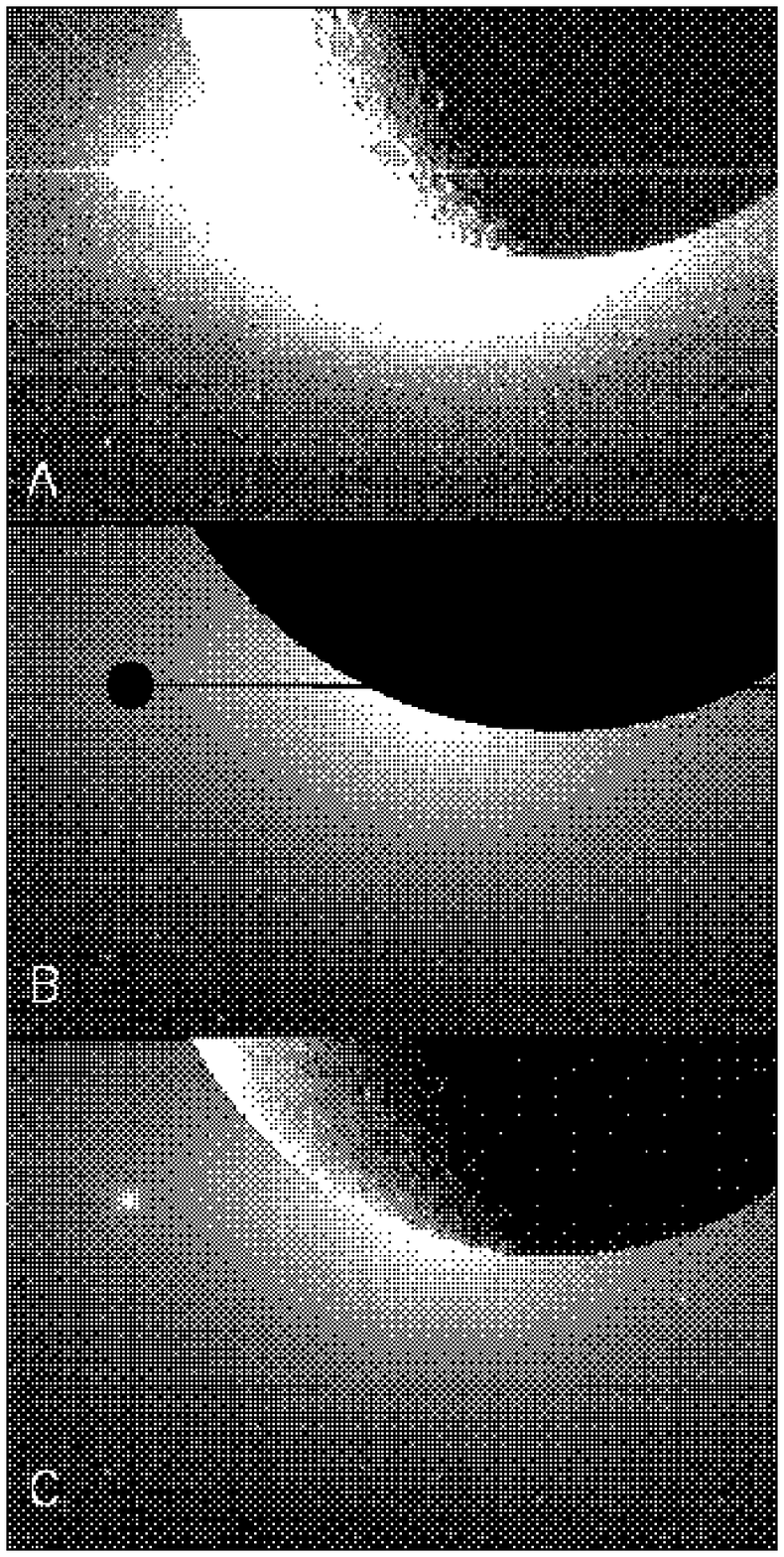}
\end{figure}

\begin{figure}
\figcaption{Star tracker images shown at different stages of
analysis. {\bf A}.\ A typical raw star tracker image
acquired during a 0.4 sec exposure in orbit 193.
We saturate this linear greyscale at a surface brightness of 
$4\times10^{-12}$ B$_\odot$
in order to reveal the faint artifacts present in the
data. The CCD's columns run left--right and ecliptic
north/east are approximately up/left. The bright object left
of the Moon is a saturated Venus whose signal has bled into the
adjacent pixels. {\bf B}.\ This logarithmic greyscale shows the
master image for orbit 193 in the surface brightness interval
$4\times10^{-13}<Z<8\times10^{-11}$ B$_\odot$. 
{\bf C}.\ This cosmetically enhanced master has had its
data--gaps filled in with images of the Moon and Venus.
\label{raw}}
\end{figure}

\begin{figure}
\epsscale{1.0}
\plotone{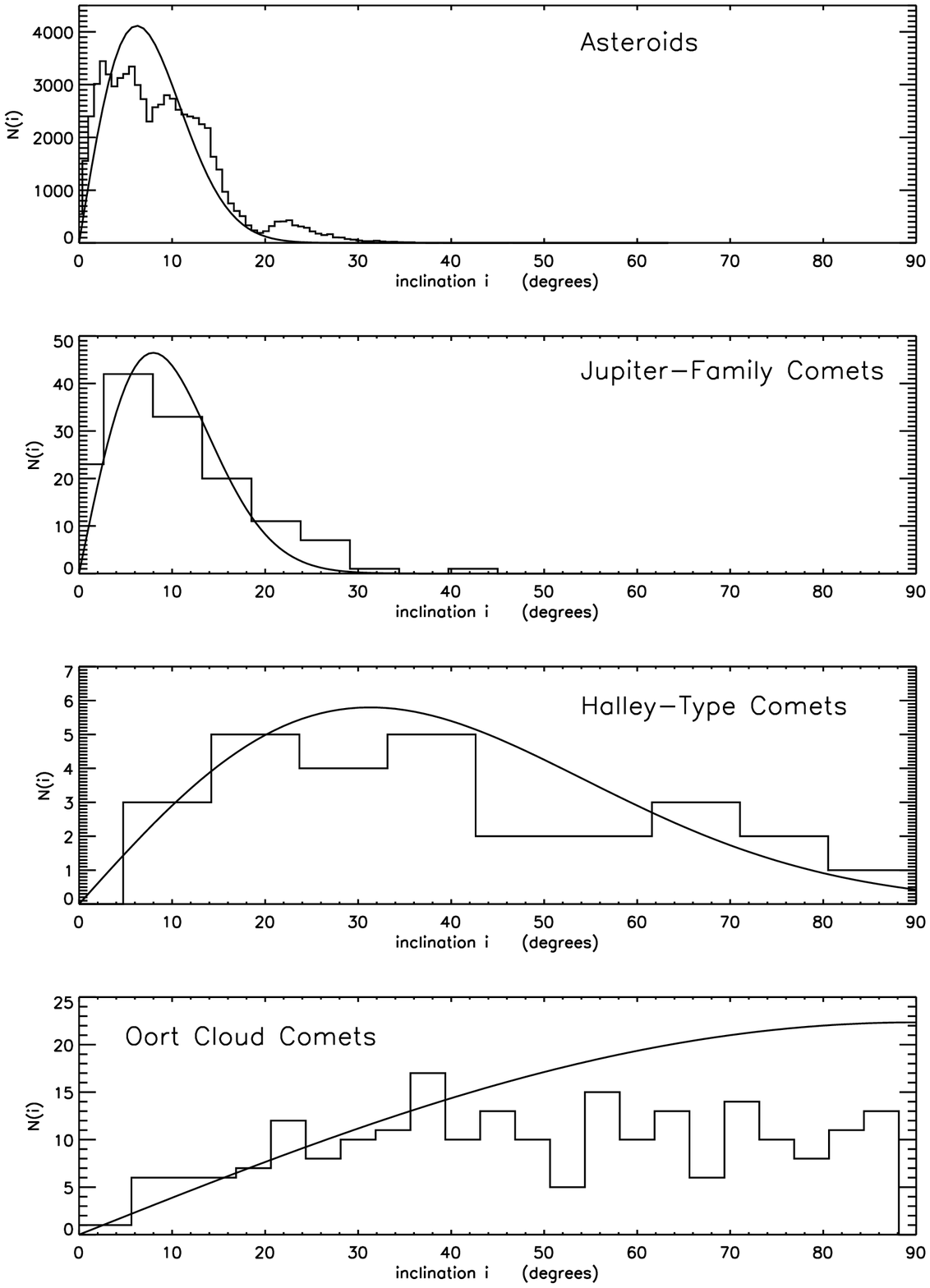}
\end{figure}

\begin{figure}
\figcaption{The upper figure shows
the inclination distribution $N(i)$
for 70,383 asteroids having semimajor axes $a\le6$ AU brighter
than absolute magnitude $H=15$ ({\it i.e}, brighter than the
completeness limit for asteroid surveys); this corresponds to
asteroids having diameters between 5 and 850 km.
These data come from Edward Bowell's
{\it The Asteroid Orbital Elements Database} obtained from the
URL ftp://ftp.lowell.edu/pub/elgb/astorb.html.
Also shown are the inclination distributions for 138
Jupiter--Family comets (JFCs), 27 Halley--Type comets (HTCs),
and 223 Oort Cloud comets (OCCs). Only those comets having
perihelia $q<2.5$ AU listed in the \cite{MW99} catalog
are used here. The smooth curves are
$g(i)\propto\sin(i)e^{-(i/\sigma)^2/2}$
with $\sigma=6.2^\circ$ for the asteroids,
$\sigma=8.0^\circ$ for the JFCs, $\sigma=33^\circ$ for the
HTCs, and $g(i)\propto\sin(i)$ for the OCCs. Since
this study of the zodiacal light is insensitive to prograde
versus retrograde orbits, we have replaced each
retrograde orbit having an inclination $i>90^\circ$ with a
prograde equivalent having an inclination $180^\circ-i$
in order to improve the statistical significance
of these figures.
\label{inc}}
\end{figure}

\clearpage

\begin{figure}
\epsscale{0.5}
\vspace*{-7ex}\plotone{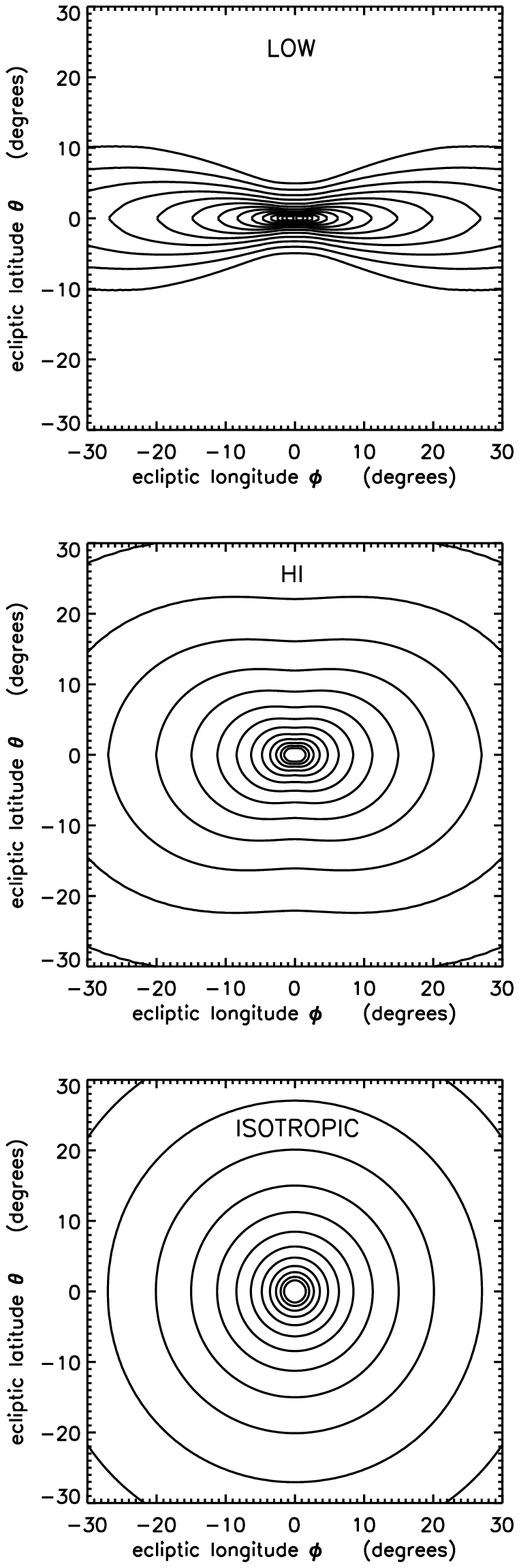}
\end{figure}

\begin{figure}
\figcaption{Isophotes for the low-inclination dust population
(e.g., dust from asteroids and JFCs), the high--inclination
population (dust from HTCs), and the isotropic population (dust
from OCCs and interstellar sources). These contours are computed
using Eq.\ \ref{Z} and the Hong phase law shown in
Fig.\ \ref{phase_fn} and with the dust
cross sectional density varying as $\sigma(r)\propto r^{-\nu_j}$
with $\nu_j=1.45$ for each population. The brightness of each
isophote differs by a factor of 2. Isophotes for populations
with $\nu_j=1$ are shaped similarly but are shifted inwards while
$\nu_j=2$ isophotes are shifted outwards.
\vspace*{5in}
\label{isophotes}}
\end{figure}

\newpage
\begin{thebibliography}{}

\bibitem[Aller {\it et al}.(1967)]{Aetal67}
Aller, L.\ H., G.\ Duffner, M.\ Dworetsky, D.\ Gudehus, S.\
Kilston, D.\ Leckrone, J.\ Montgomery, J.\ Oliver, and E.\
Zimmerman 1967. Some models of the zodiacal cloud. In {\it The
Zodiacal Light and the Interplanetary Medium} (J.\ L.\
Weinberg, Ed.), pp.\
243--256. NASA--SP 150.

\bibitem[Brown(2001)]{B01}
Brown, M.\ E.\ 2001. The inclination distribution of the Kuiper
Belt. {\it \aj}, {\bf 121}, 2804--2814.

\bibitem[Brownlee {\it et al}.(1993)]{Betal93}
Brownlee, D.\ E., D.\ J.\ Joswiak, S.\ G.\ Love,
A.\ O.\ Nier, D.\ J.\ Schlutter, and J.\ P.\ Bradley 1993.
Identification of cometary and asteroidal particles in
stratospheric IDP collections. {\it Lunar and Planetary Sci.\
XXIV}, 205.

\bibitem[Dermott {\it et al}.(1984)]{Detal84}
Dermott, S.\ F., P.\ D.\ Nicholson, J.\ A.\ Burns, and J.\ R.\
Houck 1984. Origin of the solar system dust bands discovered by
IRAS. {\it \nat}, {\bf 312}, 505--509.

\bibitem[Dermott {\it et al}.(1994)]{Detal94}
Dermott, S.\ F., D.\ D.\ Durda, B.\ \AA.\ S.\ Gustafson,
S.\ Jayaraman, J.--C.\ Liou, and Y.--L. Xu 1994.
Modern sources of dust in the solar system.
In {\it Workshop on the Analysis of Interplanetary Dust
Particles} (M.\ Zolensky, Ed.),
Lunar and Planetary Inst.\ Technical Report 94--02, pp.\ 17--18.

\bibitem[Dermott {\it et al}.(2001)]{Detal01}
Dermott, S.\ F., K.\ Grogan, D.\ D.\ Durda, S.\ Jayaraman, T.\
J.\ J.\ Kehoe, S.\ J.\ Kortenkamp, and M.\ C.\ Wyatt 2001.
Orbital evolution of interplanetary dust. In {\it Interplanetary
Dust} (E.\ Gr\"{u}n, B.\ \AA.\ S.\ Gustafson, S.\ F.\ Dermott,
and H.\ Fechtig, Eds.), Springer--Verlag, Berlin, p.\ 569.

\bibitem[Divari(1968)]{Divari68}
Divari, N.\ B.\ 1968. A meteor model for the zodiacal cloud.
{\it Soviet Physics--Astronomy}, {\bf 11}, 1048--1052.

\bibitem[Divine(1993)]{D93}
Divine, N.\ 1993. Five populations of interplanetary meteoroids.
{\it \jgr}, {\bf 98}, 17029--17048.

\bibitem[Duncan {\it et al}.(1987)]{DQT87}
Duncan, M., T.\ Quinn, and S.\ Tremaine 1987. The formation and
extent of the solar system comet cloud. {\it \aj}, {\bf 94},
1330--1338.

\bibitem[Frisch(2000)]{Frisch00}
Frisch, P.\ C.\ 2000. The galactic environment of the Sun.
{\it \jgr}, {\bf 105}, 10279--10289.

\bibitem[Giese and Dziembowski(1969)]{GD69}
Giese, R.\ H.\ and C.\ V.\ Dziembowski 1969. Suggested zodiacal
light measurements from space probes. {\it \planss}, {\bf 17},
949--956.

\bibitem[Grogan {\it et al}.(2001)]{GDD01}
Grogan, K., S.\ F.\ Dermott, and D.\ D.\ Durda 2001. The
size--frequency distribution of the zodiacal cloud: evidence
from the solar system dust bands. {\it Icarus}, {\bf 152},
251--267.

\bibitem[Gr\"{u}n {\it et al}.(1985)]{Getal85}
Gr\"{u}n, E., H.\ A.\ Zook, H.\ Fechtig, and R.\ H.\ Giese 1985.
Collisional balance of the meteoritic complex. {\it Icarus},
{\bf 62}, 244--272.

\bibitem[Gr\"{u}n {\it et al}.(1997)]{Getal97}
Gr\"{u}n, E., P.\ Staubach, M.\ Baguhl, D.\ P.\ Hamilton,
H.\ A.\ Zook, S.\ Dermott, B.\ A.\ Gustafson, H.\ Fechtig, 
J.\ Kissel, D.\ Linkert, G.\ Linkert, R.\ Srama, M.\ S.\ Hanner, 
C.\ Polanskey, M.\ Horanyi, B.\ A.\ Lindblad, I.\ Mann,
J.\ A.\ M.\ McDonnell, G.\ E.\ Morfill, and G.\ Schwehm 1997.
South--north and radial traverses through the interplanetary
dust cloud. {\it Icarus}, {\bf 129}, 270--288.

\bibitem[Hanner {\it et al}.(1974)]{Hetal74}
Hanner, M.\ S., J.\ L.\ Weinberg, L.\ M.\ Shields II, B.\ A.\
Green, and G.\ N.\ Toller 1974. Zodiacal light in the asteroid
belt: the view from Pioneer 10. {\it \jgr}, {\bf 79}, 3671--3675.

\bibitem[Hong(1985)]{Hong85}
Hong, S.\ S.\ 1985. Henyey--Greenstein representation of the
mean volume scattering phase function for zodiacal dust.
{\it \aap}, {\bf 146}, 67--75.

\bibitem[Jackson and Zook(1992)]{JZ92}
Jackson, A.\ A.\ and H.\ A.\ Zook 1992. Orbital evolution of
dust particles from comets and asteroid. {\it Icarus}, {\bf 97},
70--84.

\bibitem[Kelsall {\it et al}.(1998)]{Ketal98}
Kelsall, T., J.\ L.\ Weiland, B.\ A.\ Franz, W.\ T.\ Reach,
R.\ G.\ Arendt, E.\ Dwek, H.\ T.\ Freudenreich, M.\ G.\ Hauser,
S.\ H.\ Moseley, N.\ P.\ Odegard, R.\ F.\ Silverberg,
E.\ L.\ Wright 1998. The COBE diffuse infrared background
experiment search for the cosmic infrared background. II.
Model of the interplanetary dust cloud. {\it \apj}, {\bf 508},
44--73.

\bibitem[Kordas {\it et al}.(1995)]{Ketal95}
Kordas, J.\ F., I.\ T.\ Lewis, B.\ A.\ Wilson,
D.\ P.\ Nielsen, H.\ S.\ Park, R.\ E.\ Priest, R.\ F.\ Hills,
M.\ J.\ Shannon, A.\ G.\ Ledebuhr, and L.\ D.\ Pleasance 1995.
The star tracker stellar compass for the Clementine mission.
{\it \procspie}, {\bf 2466}, 70--83.

\bibitem[Lamy and Perrin(1986)]{LP86}
Lamy, P.\ L.\ and J.--M.\ Perrin 1986. Volume scattering
function and space distribution of the interplanetary dust
cloud. {\it \aap}, {\bf 163}, 269--286.

\bibitem[Leinert(1975)]{Leinert75}
Leinert, C. 1975. Zodiacal light---a measure of the
interplanetary environment. {\it \ssr}, {\bf 18}, 281--339.

\bibitem[Leinert {\it et al}.(1980)]{LHRP80}
Leinert, C., M.\ Hanner, I.\ Richter, and E.\ Pitz 1980.
The plane of symmetry of interplanetary dust in the inner solar
system. {\it \aap}, {\bf 82}, 328--336.

\bibitem[Leinert {\it et al}.(1981)]{Letal81}
Leinert, C., I.\ Richter, E.\ Pitz, and B.\ Planck 1981.
The zodiacal light from 1.0 to 0.3 A.U.\ as observed by the
Helios space probes. {\it \aap}, {\bf 103}, 177-188.

\bibitem[Leinert {\it et al}.(1998)]{Letal98}
Leinert, Ch., S.\ Bowyer, L.\ H.\ Haikala, M.\ S.\ Hanner,
M.\ G.\ Hauser, A.-Ch.\ Levasseur-Regourd, I.\ Mann, K.\ Mattila,
W.\ T.\ Reach, W.\ Schlosser, H.\ J.\ Staude, G.\ N.\ Toller,
J.\ L.\ Weiland, J.\ L.\ Weinberg, and A.\ N.\ Witt 1998.
The 1997 reference of diffuse night sky brightness.
{\it Astron.\ Astrophys.\ Suppl.\ Ser.}, {\bf 127}, 1-99.

\bibitem[Lester {\it et al}.(1979)]{Letal79}
Lester, T.\ P., M.\ L.\ McCall, and J.\ B.\ Tatum 1979. Theory
of planetary photometry. {\it \jrasc}, {\bf 73}, 233--257.

\bibitem[Levison and Duncan(1997)]{LD97}
Levison, H.\ F. and M.\ J.\ Duncan 1997. From the Kuiper Belt
to Jupiter--family comets: the spatial distribution of ecliptic
comets. {\it Icarus}, {\bf 127}, 13--32.

\bibitem[Levison {\it et al}.(2001)]{LDD01}
Levison, H.\ F., L.\ Dones, and M.\ J.\ Duncan 2001. The origin
of Halley--type comets: probing the inner Oort cloud. {\it \aj},
{\bf 121}, 2253--2267.

\bibitem[Lewis {\it et al}.(1991)]{Letal91}
Lewis, I.\ T., A.\ G.\ Ledebuhr, T.\ S.\ Axelrod,
J.\ F.\ Kordas, and R.\ F.\ Hills 1991. Wide--field--of--view
star tracker camera. {\it \procspie}, {\bf 1478}, 2--12.

\bibitem[Liou {\it et al}.(1995)]{Letal95}
Liou, J.\ C., S.\ F.\ Dermott, and Y.\ L.\ Xu 1995.
The contribution of cometary dust to the zodiacal cloud,
{\it \planss}, {\bf 43}, 717--722.

\bibitem[Marsden and Williams(1999)]{MW99}
Marsden, B.\ G.\ and G.\ V.\ Williams 1999. {\it Catalogue of
Cometary Orbits}, 13th ed. Central Bureau for Astronomical
Telegrams and Minor Planet Center, Smithsonian Astrophysical
Observatory, MA.

\bibitem[Neugebauer {\it et al}.(1984)]{Netal84}
Neugebauer, G., C.\ A.\ Beichman, B.\ T.\ Soifer, H.\ H.\ Aumann,
T.\ J. Chester, T.\ N.\ Gautier, F.\ C.\ Gillett, M.\ G.\ Hauser,
J.\ R.\ Houck, C.\ J.\ Lonsdale, F.\ J.\ Low, and E.\ T.\ Young
1984. Early results from the Infrared Astronomical Satellite.
{\it Science }, {\bf 224}, 14--21.

\bibitem[Reach {\it et al}.(1997)]{Retal97}
Reach, W.\ T., B.\ A.\ Franz, and J.\ L.\ Weiland 1997. The
three--dimensional structure of the zodiacal dust bands.
{\it Icarus}, {\bf 127}, 461--484.

\bibitem[Weissman(1996)]{Weissman96}
Weissman, P.\ R.\ 1996. In {\it Completing the Inventory of the
Solar System}, (T.\ W.\ Rettig and J.\ M.\ Hahn, Eds.),
pp.\ 265--288, {\it ASP Conference Series}, {\bf 107}.

\bibitem[Whipple(1955)]{W55}
Whipple, F.\ L.\ 1955. A comet model. III. The zodiacal light.
{\it \apj}, {\bf 121}, 750--770.

\bibitem[Whipple(1967)]{W67}
Whipple, F.\ L.\ 1967. On maintaining the meteoritic complex.
In {\it Zodiacal Light and the Interplanetary Medium},
(J.\ L.\ Weinberg, Ed.), NASA SP--150, pp.\ 409--425.

\bibitem[Wyatt {\it et al}.(1999)]{Wetal99}
Wyatt, M.\ C., S.\ F.\ Dermott, C.\ M.\ Telesco,
R.\ S.\ Fisher, K.\ Grogan, E.\ K.\ Holmes, and R.\ K.\ Piña
1999. How observations of circumstellar disk asymmetries can
reveal hidden planets: pericenter glow and its application
to the HR 4796 disk. {\it \apj}, {\bf 527}, 918--944.

\bibitem[Zook {\it et al}.(1997)]{Zetal97}
Zook, H.\ A., B.\ L.\ Cooper, and A.\ E.\ Potter 1997.
The zodiacal light as observed with the Clementine startracker
cameras: calibration and image analysis procedures.
{\it Lunar and Planetary Science XXVIII}, abstract \#1103.

\end{thebibliography}
\end{document}